\documentclass[a4paper,11pt]{article}
\pdfoutput=1
\usepackage{jcappub}

\usepackage{amsmath,amssymb,amsfonts,latexsym,mathrsfs}
\usepackage[utf8]{inputenc}
\usepackage{epsfig, graphicx, verbatim,xspace, multirow, mathtools}
\usepackage{booktabs}
\usepackage{pifont}
\usepackage{url}
\usepackage{bm}
\usepackage{array}
\usepackage{enumitem}
\usepackage[normalem]{ulem}
\usepackage{xcolor}

\newcommand{\nunui}{\nu_{i}\bar{\nu}_i}

\newcommand{\nunuL}{\nu_L\bar{\nu}_L}
\newcommand{\nunuiL}{\nu_{L}^{i}\bar{\nu}_L^i}
\newcommand{\nunuR}{\nu_R\bar{\nu}_R}
\newcommand{\nunuiR}{\nu_{R}^{i}\bar{\nu}_R^i}

\newcommand{\epm}{e^{\pm}}
\newcommand{\ee}{e^{+}e^{-}}
\newcommand{\aab}{a\bar{a}}
\newcommand{\lra}{\leftrightarrow}

\newcommand{\smT}{\text{SM}}

\newcommand{\nuL}{{\nu_{L}}}
\newcommand{\nuR}{{\nu_{R}}}

\newcommand{\neff}{N_{\text{eff}}}
\newcommand{\dneff}{\Delta N_{\text{eff}}}

\newcommand{\mev}{\text{MeV}}
\newcommand{\kev}{\text{keV}}

\newcommand{\LiF}{\,\text{Li}_4}

\newcommand{\cmb}{^\text{CMB}}
\newcommand{\bbn}{^\text{BBN}}

\newcommand{\ra}{\rightarrow}

\begin{document}

\title{\boldmath Cosmological implications of gauged $U(1)_{B-L}$ on $\Delta N_{\rm eff}$ in the CMB and BBN}

 \author{Haidar Esseili}
 \author{and  Graham D. Kribs}
 \affiliation{Department of Physics and Institute for Fundamental Science, \\ University of Oregon, Eugene, Oregon 97403, USA}

\emailAdd{hesseili@uoregon.edu}
\emailAdd{kribs@uoregon.edu}

\abstract{  We calculate the effects of a light, very weakly-coupled boson $X$  
  arising from a spontaneously broken $U(1)_{B-L}$ symmetry
  on $\Delta N_{\rm eff}$ as measured by the CMB and
  $Y_p$ from BBN\@.
  Our focus is the mass range
  $1 \; {\rm eV} \, \lesssim m_X \lesssim 100 \; {\rm MeV}$;
  masses lighter than about an ${\rm eV}$ have strong constraints from
  fifth-force law constraints, while masses heavier than about
  $100$~MeV are constrained by other probes, including terrestrial
  experiments.
  We do \emph{not} assume $X$ began in thermal equilibrium with the SM;
  instead, we allow $X$ to freeze-in from its very weak interactions
  with the SM\@.  We find $U(1)_{B-L}$ is more strongly constrained 
  by $\Delta N_{\rm eff}$ than previously considered.  The bounds
  arise from the energy density in electrons and neutrinos slowly
  siphoned off into $X$ bosons, which become nonrelativistic,
  redshift as matter, and then decay, dumping their slightly larger
  energy density back into the SM bath causing $\Delta N_{\rm eff} > 0$.
  While some of the parameter space has complementary
  constraints from stellar cooling, supernova emission, and
  terrestrial experiments, we find future CMB observatories
  including Simons Observatory and CMB-S4 can access regions of mass
  and coupling space not probed by any other method.
  In gauging $U(1)_{B-L}$, we assume the $[U(1)_{B-L}]^3$
  anomaly is canceled by right-handed
  neutrinos, and so our $\Delta N_{\rm eff}$ calculations have been
  carried out in two scenarios:  neutrinos have Dirac masses, or,
  right-handed neutrinos acquire Majorana masses.
  In the latter scenario, we comment on the additional implications
  of thermalized right-handed neutrinos decaying during BBN\@. 
  We also briefly consider the possibility that $X$ decays
  into dark sector states.  If these states behave as radiation,
  we find weaker constraints, whereas if they are massive,
  there are stronger constraints, though now from
  $\Delta N_{\rm eff} < 0$.}



\maketitle
\flushbottom

\setcounter{page}{2}


\newpage

\section{Introduction}
\label{sec:Introduction}

Light mediators -- new massive, unstable particles that couple
to the SM -- have seen tremendous interest over the past decade
(for reviews, see \cite{Alexander:2016aln,Battaglieri:2017aum}).
They are essential to models of
light dark matter \cite{Arkani-Hamed:2008hhe,Pospelov:2008jd,Pospelov:2008zw},
providing mechanisms that lead to the correct abundance
as well as providing detectable signals in the myriad landscape
of light dark matter detection experiments.  A panoply of
experiments have been considered to gain sensitivity to these
mediators.  

Specializing to light $U(1)$ vector boson mediators, several varieties
have been considered:  the dark photon (for a recent review, see
\cite{Fabbrichesi:2020wbt}); 
$U(1)_{B-L}$
(early discussions include
\cite{Davidson:1978pm, Marshak:1979fm,Mohapatra:1980qe,Wetterich:1981bx}
and a review of massive $Z'$s \cite{Langacker:2008yv}),
while gauged $U(1)_{B-L}$ as a light mediator has been considered
in detail in \cite{Harnik:2012ni,Heeck:2014zfa,Bauer:2018onh,Ilten:2018crw};
$U(1)_B$ \cite{Nelson:1989fx,Carone:1994aa,FileviezPerez:2010gw};
and flavor-dependent
$U(1)$s such as $U(1)_{\mu - \tau}$
\cite{Foot:1990mn,He:1990pn,He:1991qd}
(some recent work \cite{Escudero:2019gzq,Dror:2020fbh}).
A huge variety of constraints
restrict the parameter space of these light vector bosons
\cite{Harnik:2012ni,Heeck:2014zfa,Bauer:2018onh,Ilten:2018crw,Okada:2020cue}
including colliders, fixed target experiments, energy loss from stars;
energy loss from supernovae, changes to big-bang nucleosynthesis; etc.

Our focus in this paper is to calculate the contributions
to the effective number of relativistic species, $N_{\rm eff}$,
from a light $U(1)_{B-L}$ gauge boson mediator $X$, 
determining both the existing constraints and prospects for future
CMB observatories.
Precision determination of $\neff$, from CMB power spectra \cite{Planck:2018nkj, Planck:2018vyg} and primordial element abundances \cite{Cyburt:2015mya, Yeh:2020mgl, Yeh:2022heq, Workman:2022ynf} is in strong agreement with SM prediction \cite{deSalas:2016ztq, Mangano:2005cc, Dolgov:2002wy, Dicus:1982bz, Hannestad:1995rs, Dodelson:1992km, Dolgov:1997mb, Esposito:2000hi, Mangano:2001iu, Birrell:2014uka, Grohs:2015tfy, Cielo:2023bqp, Akita:2020szl, Froustey:2020mcq, Bennett:2020zkv}.
This makes $\neff$ a powerful tool in testing BSM physics that affects early universe cosmology before recombination \cite{Sarkar:1995dd, Iocco:2008va, Pospelov:2010hj, Blennow:2012de, Boehm:2012gr, Boehm:2013jpa, Brust:2013ova, Vogel:2013raa, Fradette:2014sza, Nollett:2014lwa, Buen-Abad:2015ova, Chacko:2015noa, Wilkinson:2016gsy, Huang:2017egl, Escudero:2018mvt, Escudero:2019gzq, Abazajian:2019oqj, Ibe:2019gpv, EscuderoAbenza:2020cmq, Coffey:2020oir, Luo:2020sho, Luo:2020fdt, Adshead:2022ovo, Eijima:2022dec, Sandner:2023ptm}. 
We carry out the full Boltzmann evolution, allowing
the $X$ boson to ``freeze in'' \cite{Hall:2009bx},
become nonrelativistic, redshift as matter, and then decay back into
SM states, modifying $\Delta N_{\rm eff}$.
Earlier calculations have considered the contributions to
$\Delta N_{\rm eff}$ from the freeze in of a
dark photon \cite{Ibe:2019gpv} and $U(1)_{\mu-\tau}$ \cite{Escudero:2019gzq}.
The cosmological constraints from $\Delta N_{\rm eff}$ on the dark photon
are relatively strong when the dark photon mass is near the
temperature of BBN, $\sim 1$~MeV, when the dark photon
can interact with the electron plasma and thus (indirectly)
affect the neutrino energy density, but weaker below this
since the dark photon does not interact with neutrinos.
By contrast, mediators that freeze in from their interactions with
neutrinos have substantially stronger constraints,
e.g., $U(1)_{\mu-\tau}$ \cite{Escudero:2019gzq}.
In this paper, we utilize the formalism and approximations outlined in
\cite{Escudero:2018mvt,Escudero:2019gzq,EscuderoAbenza:2020cmq}, 
extended and applied to the case of a $U(1)_{B-L}$ gauge boson. 

The central observation is that an out-of-equilibrium abundance
of new particles that interact with the SM can be well approximated
by using \emph{equilibrium} distributions with nonzero chemical
potentials.  We will discuss the formalism and approximations in detail,
as applied to $U(1)_{B-L}$.  The result is that since $U(1)_{B-L}$
has a nonzero coupling to neutrinos, there are strong bounds from
$\Delta N_{\rm eff}$ from the CMB that get \emph{stronger}
with \emph{decreasing} mass of $X$, down to of order $m_X \sim 1$~eV\@.
In the regime $m_X < 2 m_e$, our results are qualitatively consistent with
a similar analysis done for $U(1)_{\mu - \tau}$ \cite{Escudero:2019gzq}. 
At even smaller masses, $m_X \lesssim 1$~eV, fifth force constraints
\cite{Bordag:2001qi,Bordag:2009zz,Adelberger:2006dh,Adelberger:2009zz}
become very strong and dominate the bounds \cite{Harnik:2012ni}.
For gauge boson masses near (and below) $1$~eV, a more sophisticated
treatment of the CMB is necessary to fully elucidate cosmological bounds.
We will comment on this in the Discussion.  

To our knowledge,
the first paper that considered constraints on
the mass and coupling of a light $U(1)_{B-L}$ gauge boson
from its effects on $\Delta N_{\rm eff}$ is
\cite{Knapen:2017xzo}.  They considered a light 
$U(1)_{B-L}$ gauge boson with just left-handed neutrinos
among the fermionic relativistic degrees of freedom.
The critical difference between our study and \cite{Knapen:2017xzo}
is that the latter only considered the constraints
from $\Delta N_{\rm eff}$ from BBN, 
using simple thermalization scaling arguments.
Ref.~\cite{Knapen:2017xzo} found that 
for $m_X \gtrsim 10$~MeV, requiring the scattering
process $e^- \nu \ra e^- \nu$ through off-shell 
$X$ exchange is not larger than the weak interaction
sets a constraint.  We verify this constraint
also applies to $\Delta N_{\rm eff}$ measured by the CMB,
using our Boltzmann equation evolution.
Ref.~\cite{Knapen:2017xzo}
also considered $m_X < 10$~MeV\@.  In this region,
\cite{Knapen:2017xzo}
required only that $X \leftrightarrow \nu\bar{\nu}$ did not reach
thermal equilibrium at $T \sim 1$~MeV, so that $X$ does not
contribute excessively to $\Delta N_{\rm eff}$ during BBN\@.
As we will see, we find much stronger constraints
in the region $m_X < 10$~MeV from detailed numerical
calculations of $\Delta N_{\rm eff}$ at the CMB era
as well as strong constraints on
the helium mass fraction $Y_p$, at the BBN era.

There are also constraints on \emph{very} weakly coupled
mediators that are completely out-of-equilibrium,
but decay on timescales that can disrupt BBN
\cite{Kawasaki:2004qu,Kanzaki:2007pd,Kawasaki:2020qxm}
or to cause
spectral distortion in the CMB 
\cite{Fradette:2014sza,Berger:2016vxi, Coffey:2020oir}.
We show these constraints on the parameter space of 
$U(1)_{B-L}$, obtained from \cite{Coffey:2020oir}, 
that are complementary to our results.

\section[Gauged U(1)\texorpdfstring{$_{B-L}$}{B-L}:  Majorana and Dirac neutrino cases]{Gauged U(1)$_{B-L}$:  Majorana and Dirac neutrino cases}

$U(1)_{B-L}$ is among the most interesting possible new forces since
it is the only flavor-universal global symmetry of the SM
that is gaugeable.
All mixed $U(1)_{B-L} \times {\rm SM}^2$ anomalies automatically
vanish within the SM\@.  The $[U(1)_{B-L}]^3$ anomaly
remains, requiring additional chiral fermions that transform under
just $U(1)_{B-L}$.  The simplest solution is to add one chiral fermion per
generation with lepton number equal and opposite to $\nu_L$, namely,
one right-handed neutrino per generation.

To establish notation, gauged $U(1)_{B-L}$ is mediated by a vector
boson $X^\mu$ that interacts with the SM quarks and leptons
with charge
\begin{equation}
  q_f = (B - L)_f =
  \left\{ \begin{array}{lcl} +1/3 &\quad& Q_L, \; u_R, \; d_R  \\
                             -1 &\quad& L, \; e_R, \; \nu_R 
          \end{array} \right. \, ,
\end{equation}
where we have written the fermions in four-component notation
consistent with \cite{Peskin:1995ev}. 
We assume $U(1)_{B-L}$ is broken, and so $X$ acquires a mass $m_X$.
The spontaneous breaking of $U(1)_{B-L}$ can be accomplished by 
introducing a complex scalar, $\phi_X$,
transforming under $B-L$ with charge $q_X$,
and gauge coupling $g_X$, with a potential engineered to
spontaneously break the symmetry.  The scalar Lagrangian is
\begin{equation}
  \mathcal{L} \; = \; ( D_\mu \phi_X )^\dagger D^\mu \phi_X
  - V(\phi_X)
\end{equation}
where $D^\mu \equiv \partial^\mu - i g_X q_X X^\mu$, with
\begin{equation}
  V(\phi_X) \; = \; \lambda_X
  \left( \phi_X^\dagger \phi_X - \frac{v_X^2}{2} \right)^2
\end{equation}
and the minimum of the scalar potential occurs at
$\langle \phi_X \rangle = v_X/\sqrt{2}$. 
Expanding around the minimum, with $\phi_X = (h_X + v_X)/\sqrt{2}$,
where $h_X$ is the $B-L$ Higgs boson,
the physical states in the theory have masses
\begin{eqnarray}
X^\mu: &\quad& m_X \, = \, g_X q_X v_X \\
  h_X: &\quad& m_{h_X}^2 \, = \, 2 \lambda_X v_X^2  \; .
\end{eqnarray}
Throughout the paper, we work in the limit
$g_X q_X \ll \lambda_X$, and so, $m_X \ll m_{h_X}$.
Consequently, we will not need to consider the $h_X$
participating in the degrees of freedom in the thermal bath
for $\Delta N_{\rm eff}$ calculations.
The ability to adjust the $B-L$ charge of $\phi_X$ implies 
$m_X$ can be made arbitrarily small relative to $g_X v_X$.
That is, from the perspective of the gauge and Higgs sector,
one can independently adjust $m_X$ and $g_X$ (as well as $m_{h_X}$,
through $\lambda_X$). 

\subsection{Neutrino masses:  Majorana and Dirac cases}

We do need to address neutrino masses.  Given that $U(1)_{B-L}$ is gauged,
the usual dimension-5 Weinberg operator for Majorana neutrino masses,
$(L H)^2/\Lambda$, is forbidden.  Instead, Yukawa couplings of 
left-handed and right-handed neutrinos are permitted,
\begin{equation}
  y_{D}^{ij} \bar{L}_i H^\dagger \nu_{R,j} + h.c. \, ,
  \label{eq:dirac}
\end{equation}
giving neutrinos Dirac masses that preserve $U(1)_{B-L}$.
This ``\textbf{Dirac case}'' is one of the two cases we will consider in
this paper.  In the Dirac case, since neutrinos acquire their mass
from Yukawa couplings to the SM Higgs field, there is (still) no
restriction on the ability to adjust
$m_X$ and $g_X$ independently.

The alternative, ``\textbf{Majorana case}'', is one where
the right-handed neutrinos that are required to cancel
the $[U(1)_{B-L}]^3$ anomaly acquire
Majorana masses after $U(1)_{B-L}$ is spontaneously broken.
When combined with the Dirac mass terms, equation~(\ref{eq:dirac}),
after electroweak symmetry is broken, 
this leads to the usual see-saw formula that results in 
left-handed neutrinos acquiring small Majorana masses.
Right-handed neutrinos can acquire mass with just one
Higgs field transforming under $U(1)_{B-L}$ if 
the $B-L$ charge is fixed to be $q_X = \pm 2$, such that
Yukawa-like interactions are permitted, 
\begin{equation}
  \frac{y_{M}^{ij}}{2} \nu_{R,i}^c \nu_{R,j}^c \phi_X^{(\dagger)} + h.c. \, ,
\end{equation}
where $\nu_{R,i}^c$ are the right-handed neutrinos written with
2-component left-handed fermion notation (in order to avoid Majorana
notation). 
In this scenario, the vev of $\phi_X$ not only gives mass
to the gauge boson, but it also gives Majorana masses to
the right-handed neutrinos, $M_R \sim y_M v_X$.

For us, the key distinction is that in the Majorana case,
the right-handed neutrinos can be much heavier, and thus
not contribute in any way to $\Delta N_{\rm eff}$.
This is in contrast to the Dirac scenario, where
if the right-handed neutrinos were ever in equilibrium
(through, for example, $X^\mu$ exchange), they necessarily
contribute to $\Delta N_{\rm eff}$
\cite{Abazajian:2019eic, Abazajian:2019oqj,Luo:2020sho}. 
The size of the contribution to $\Delta N_{\rm eff}$
is controlled just by the dilution of the number of relativistic
degrees of freedom after heavier SM fields annihilate (or decay)
and dump their entropy into the photon bath.

It is interesting to consider the bounds on the Majorana masses
for right-handed neutrinos from cosmology.  Obviously if the
right-handed neutrinos were in thermal equilibrium with the SM,
they would excessively contribute to $\Delta N_{\rm eff}$
during BBN and CMB if their mass were less than approximately
$10$~MeV\@.  Thermal equilibrium is naturally achieved through
$X^\mu$ exchange, so long as $g_X$ is not excessively small.
(We'll quantify this in detail later in the paper.)
Once the right-handed neutrinos are heavier than about
$10$~MeV, their abundance would be at least somewhat suppressed
as they become nonrelativistic once the temperature of the Universe
drops well below their mass.

Even if right-handed neutrinos decoupled early in the Universe,
they can and will decay to left-handed neutrinos
and a (possibly off-shell) Higgs boson,
with a significant suppression in the rate due to the smallness
of the (Dirac) Yukawa coupling of the right-handed
neutrinos to the left-handed neutrinos.  In
appendix~\ref{app:righthandedneutrinodecay},
we estimate the rate, and find that for $M_R \gtrsim 20$~GeV,
the right-handed neutrinos decay before the onset of BBN\@.

Finally, it is also instructive to consider the case where
$U(1)_{B-L}$ is explicitly broken without any scalar sector,
i.e., a St\"uckelberg mass for the $X$ vector field
(for a recent detailed discussion of the St\"uckelberg mechanism,
see \cite{Kribs:2022gri}). 
In the Dirac scenario, we can stop there, since Dirac masses respect
$U(1)_{B-L}$ (see also \cite{Heeck:2014zfa}).  (This is equivalent,
in the spontaneously broken theory,
to holding $m_X = q_X g_X v_X$ fixed, $\lambda_X$ fixed,
but then taking $q_X \ll 1$ while $v_X$ is taken large.
This limit permits any (perturbative) value for $g_X$.)
In the Majorana case, we must also explicitly break
$U(1)_{B-L}$ by $2$ units when writing explicit Majorana masses
$(M_R)_{ij} \nu^c_i \nu^c_j + h.c.$.  The explicit breaking
implies the $[U(1)_{B-L}]^3$ anomaly will appear below
the scale $M_R$.  Thus, the right-handed neutrinos induce a
3-loop anomalous contribution to the mass of the
(St\"uckelberg) $X^\mu$ vector field.  The size of this
contribution is easily estimated \cite{Preskill:1990fr,Craig:2019zkf}
\begin{eqnarray}
\Delta m_X \simeq \frac{g_X^3 M_R}{64 \pi^3}
\end{eqnarray}
which implies a lower bound on the mass of $X^\mu$
that decreases rapidly as $g_X$ is lowered.  
This bound is always weaker than equation~(\ref{eq:gXbound}),
so that if we require right-handed neutrinos decay before BBN,
there is no further constraint.  Only if $M_R$ were large,
such as a traditional see-saw mechanism with $y_D \sim 1$,
with $M_R \sim 10^{14-15}$~GeV, would there be any
constraint at all on $g_X$, though even then this constraint
is quite mild.


\section[Effective number of relativistic species: \texorpdfstring{$\neff$}{Neff}]{Effective number of relativistic species: $\neff$}
\label{neffExplanationSec}

In standard cosmology at temperature
$T \sim 10$~MeV \cite{Weinberg:2008zzc}, electrons, photons, and neutrinos  were in thermal equilibrium. As the universe cooled, the weak interaction rate dropped
below the Hubble expansion rate and neutrinos decoupled from the electromagnetic plasma around $T\sim 2$~MeV\@.
As the universe continued to cool, the temperature dropped below the
electron mass, $\ee$ annihilation depleted the vast majority of
charged leptons.  In the limit of instantaneous neutrino decoupling, the electron-positron entropy was transferred solely to photons resulting in a temperature ratio of $T_\gamma/T_\nu=(11/4)^{1/3} \approx 1.401$ after annihilation completed.
However, the weak interaction remained slightly active during $\ee$ annihilation, resulting in a small but appreciable heating of neutrinos. 
This gives rise to a small increase to the energy density of neutrinos,
conventionally defined by the \textit{effective number of relativistic species},
 \begin{equation}
\label{NeffEqn}
\neff \;=\; \frac{8}{7}
\left( \frac{11}{4}  \right)^{4/3}
\left( \frac{\rho_{\text{rad}}-\rho_\gamma}{\rho_\gamma} \right),
\end{equation}
which is the ratio of non-photon to photon radiation density. Any new BSM radiation density present well before recombination can be treated as an additional contribution to $\neff$. The normalization in equation~(\ref{NeffEqn}) is chosen such that $\neff=N_\nu=3$ in the instantaneous neutrino decoupling limit. 

The state-of-the-art calculation in the SM gives
$\neff^{\text{SM}} = 3.045$-$3.046$, 
that takes into higher order corrections, non-thermal neutrino
distribution functions, neutrino oscillations,
etc.~\cite{Mangano:2005cc, deSalas:2016ztq}. 
In this paper, we have followed
Refs.~\cite{Escudero:2018mvt, EscuderoAbenza:2020cmq}
that have provided an efficient calculation of $N_{\rm eff}$ which
employs certain
approximations that nevertheless result in excellent accuracy,
which we will discuss in detail in section~\ref{approxSec}. 
For instance, using this method, we obtain
the photon-to-neutrino temperature ratio, the neutrino chemical
potential, and $N_{\rm eff}$ in the SM
 \begin{equation}
\label{NeffinSM}
T_\gamma/T_\nu = 1.3945 \, ,
\quad \quad \quad 
\mu_\nu/T_\nu = - 4.82\times 10^{-3} \, ,
\quad \quad
\neff^{\text{SM}} = 3.042 \, . 
\end{equation}
The point of re-doing the SM calculation here is to demonstrate that we
can achieve reasonable accuracy of 
$N_{\rm eff}$ even with the approximations that have been employed. 
The very small $\sim 0.1\%$ discrepancy between our calculation and
the precise determination is slightly accidental -- some of the
effects we have neglected, that contribute at a level of $\sim 0.3\%$,
happen to very nearly cancel out when summed together
(see \cite{EscuderoAbenza:2020cmq} for details).
In any case, our calculation is able to reproduce the
non-instantaneous decoupling of neutrinos in the SM to an accuracy
of order $\Delta N_{\rm eff} \sim 0.01$.

After $\ee$ annihilation is complete, the number of relativistic degrees
of freedom remains the same in the SM down to the
CMB era.\footnote{We do not need to consider SM neutrino masses,
  since they are bounded to be smaller than the temperatures
  we consider in the paper \cite{Planck:2018vyg}.}
In the presence of physics
beyond the SM, there can be new degrees of freedom that
appear (or disappear) before or after BBN\@.  This can be 
characterized by the number of relativistic degrees of freedom
at CMB, $N_{\rm eff}^{\rm CMB}$, to be distinguished from the
number of relativistic degrees of freedom at BBN,
$N_{\rm eff}^{\rm BBN}$.  In this paper, $\neff$ refers
exclusively to $N_{\rm eff}^{\rm CMB}$,
and hereafter $\neff$ and $\dneff$ refer to the quantities
at the CMB era.  We do, however, calculate the shift to
the helium mass fraction $Y_p$ at BBN, separately from $\dneff$. 
We will present our calculations of $\neff$
in section~\ref{sec:numericalevolutionneff}
and $Y_p$ in section~\ref{bbnSec}, and compare to the observational
determinations at the end of each of those sections.


\section{Early universe thermodynamics}
\label{sec:thermo}

Our method to calculate thermodynamic quantities in the early universe
utilizes several approximations in order to solve the Boltzmann equations
that were described in detail in
\cite{Escudero:2018mvt, EscuderoAbenza:2020cmq}.\footnote{We have
  benefited from viewing the code 
  \href{https://github.com/MiguelEA/nudec_BSM}{NUDEC\_BSM}
  as a reference
  to setup our calculations.  However, all of our calculations are
  based on our own code.}
The key insight from \cite{Escudero:2018mvt, EscuderoAbenza:2020cmq}
is that we can approximate the effects of out-of-equilibrium
(``freeze-in'') $X$ bosons using \emph{equilibrium} distributions
with nonzero chemical potentials.
How this works requires some explanation.
At temperatures near BBN, the dominant contributions to
the energy density are from electrons (and positions), photons,
and neutrinos.  In the SM, the annihilation and scattering rates
between electrons, positions, and photons is very efficient
ensuring $T_\gamma = T_e$.  
Since photon number is not conserved, the processes
$\ee \leftrightarrow \gamma$, $\ee \leftrightarrow \gamma\gamma$,
and $\ee \leftrightarrow \gamma\gamma\gamma$ imply that
the chemical potential for photons, electrons and positions 
vanishes, $\mu_\gamma = \mu_e = 0$ (to a very good approximation
\cite{Thomas:2019ran}).  When neutrino-electron scattering
is active for temperatures $T \gtrsim 2$~MeV, the processes
$e^{+}e^{-}\lra\bar{\nu}_\alpha\nu_\alpha$,
$e^{\pm}\nu_\alpha \lra e^{\pm}\nu_\alpha$, and
$e^{\pm}\bar{\nu}_\alpha \lra e^{\pm}\bar{\nu}_\alpha$
ensure that $\mu_\nu \simeq 0$.  
Hence, in the SM, chemical potentials do not play a critical role
in determining the thermodynamic evolution near and below
the BBN era.

When we introduce a light $U(1)_{B-L}$ gauge boson $X$, there are
three possible regimes of interest:
heavy ($m_X \gg 1$~MeV), intermediate ($m_X \sim 1$~MeV),
and light ($m_X \ll 1$~MeV).  In all regimes, we assume
$\mu_\gamma = \mu_e = 0$ throughout the Boltzmann evolution,
given that the electromagnetic interactions will always be much
faster than interactions among $X$ bosons.  Nevertheless, we allow
for chemical potentials for $X$ and neutrinos to develop and evolve
with temperature, as $X$ freezes-in through its very weak
interactions with the SM\@.  As we will see, the thermodynamic
evolution of the SM particles plus $X$ will be quite different in 
these different regimes.  

In the heavy regime, $m_X \gg 1$~MeV,
as the temperature drops below $m_X$, 
the $X$ bosons become nonrelativistic while the weak
interactions that keep electrons and neutrinos in thermal and
chemical equilibrium remain active.
At these high temperatures, there is competition between $1 \lra 2$
processes involving $X \lra \nu\bar{\nu}$,
that will cause $\mu_X$ and $\mu_\nu$ to develop,
and the electroweak-mediated $2 \lra 2$ processes,
that drive $\mu_\nu \ra 0$.  Initially, $X$ is
out-of-equilibrium, and so a chemical potential for
$X$ and neutrinos can (and will) develop.
As the temperatures decrease, $X$ becomes nonrelativistic,
the $2 \lra 2$ processes dominate, driving
$\mu_\nu$ (and $\mu_X$) to small values.
At still larger masses, for temperatures
$10 \; {\rm MeV} \lesssim T \ll m_X$,
the $1 \lra 2$ processes are irrelevant,
and instead off-shell $X$-exchange can contribute to 
$2 \leftrightarrow 2$ processes qualitatively similar
to electroweak gauge boson exchange.  Here, there is a
constraint on the strength of the $X$ boson
interactions with the SM that arises from
delaying neutrino freeze-out, but this is much
weaker than the constraints from lighter masses,
as we will see below.

In the light regime, when $m_X \ll 1$~MeV, electron-position
annihilation is fully complete, leaving only photons,
neutrinos, and $X$ as the relativistic degrees of freedom.
In this regime, neutrinos are out of thermal and chemical
equilibrium with the SM, and thus as $X$ freezes-in,
the evolution of Boltzmann equations result in a
chemical potential for $X$ and neutrinos
through the process is $X \lra \nu\bar{\nu}$.
If this is efficient enough to reach chemical equilibrium,
$\mu_X = 2 \mu_\nu$.  Note that since $U(1)_{\rm B-L}$ is
flavor-conserving, we necessarily have
$\mu_\nu = \mu_{\bar{\nu}}$.

Finally, the intermediate regime, $m_X \sim 1$~MeV,
is the trickiest one to model.
When $T \sim 1$~MeV, the weak interactions have recently
decoupled, and so the $2 \lra 2$ electroweak processes that enforce
$\mu_\nu \simeq 0$ have just recently shut off.
This means that as a chemical potential for $\mu_X$
develops from its out-of-equilibrium production,
the resulting $\mu_\nu$ that also develops, can remain. 
However, electron-photon interactions are in thermal and
chemical equilibrium, and so $\mu_e = 0$.  If $X$ were to be
in thermal equilibrium with both electrons and neutrinos,
the electron interactions $X \lra \ee$ would bias $\mu_X \simeq 0$.
Instead, when $X$ is out-of-equilibrium, a chemical potential
for $X$ can develop as $X$ freezes-in from both
$\ee \ra X$ and $\nu\bar{\nu} \ra X$ interactions. 
As the universe cools, more $X$ is produced, but then
$\ee$ annihilation into photons rapidly depletes the
electron-positron bath.  This means $X$ could reach thermal equilibrium
with neutrinos, with a nonzero chemical potential, since 
the remaining electrons and positions have dropped out of
chemical equilibrium.

\subsection{Approximations}
\label{approxSec}

We assume that all fermions and bosons follow Fermi-Dirac (FD, positive) and Bose-Einstein (BE, negative), $f(E)=[e^{(E-\mu)/T}\pm1]^{-1}$ distribution functions respectively. This assumption is well-established when the energy and momentum exchange between particles is efficient. If interactions are not fully efficient, i.e.\ out-of-equilibrium evolution, distributions may obtain spectral distortion corrections. An example of this is shown in figures $9,10$ of \cite{Escudero:2019gzq} for the case of $U(1)_{L_\mu-L_\tau}$. These corrections are expected to be small.

We assume, to an excellent approximation, that the electron/positron plasma is highly thermalized with the photon plasma such that $T_\gamma=T_e$ and $\mu_\gamma=\mu_e=0$. Here, we leverage the strong annihilation and scattering rate between electrons, positrons, and photons and that the number of photons is not conserved in the early universe. As for neutrinos, we describe a neutrino fluid with a single distribution characterized by $T_\nu \equiv T_{\nu_e}=T_{\nu_\mu}=T_{\nu_\tau}$ and $\mu_\nu$. Here, we ignored neutrino oscillations which become active for temperatures  $T=3$-$5$~MeV, \cite{Hannestad:2001iy,Dolgov:2002ab}, and model this effect by setting the temperature of the different neutrino species equal. The correction due to neutrino oscillations in SM is $\dneff=0.0007$ \cite{deSalas:2016ztq}. The correction due to evolving distinct neutrino species temperature rather than a single $T_\nu$ was found to be $\dneff=0.001$ using the approximations \cite{Escudero:2018mvt, EscuderoAbenza:2020cmq} we employ in this paper.

For the collision terms we have implemented the correct
particle distributions for the $1 \lra 2$ collision terms, 
and approximate particle distributions for 
$2 \lra 2$ collision terms.  For the $1 \lra 2$ processes, 
this means we use a Fermi-Dirac distribution for fermions,
a Bose-Einstein distribution for bosons, and include 
Pauli blocking and Bose enhancement.
For the $2 \lra 2$ processes, we use Maxwell-Boltzmann distributions
for all particles, 
which reduces the number of integrations needed for the collision terms
and significantly decreases the numerical computation time.
Our implementation of using the correct statistics for the
$1 \lra 2$ collision processes is motivated by the relative importance
of these processes in determining an accurate calculation of $\neff$.
While the qualitative features of our results remain unaffected
even if Maxwell-Boltzmann distributions were used for
$1 \lra 2$ processes, quantitatively we find that the contours
of $\dneff$ shift to slightly overestimate the impact of the
$X$ boson within the $(m_X, g_X)$ parameter space.

Finally, we have also taken into account finite temperature effects that result from shifts in the masses and self-energies from their vacuum values.  
In the SM, these finite temperature effects \cite{Mangano:2001iu, Fornengo:1997wa, Escudero:2018mvt} result in an $\mathcal{O}(0.01)$ correction to $\neff$. We have outlined the relevant finite temperature effects in appendix \ref{app:finiteTemperatureEffects}, where we show that for the case $m_X>2m_e$, these corrections yield at most $1-2 \%$ shift relative to the vacuum evolution, and so we neglect them in this regime.  This result is  consistent with \cite{Redondo:2008ec} where finite temperature corrections to the dark photon in early universe were found to be negligible in this mass range. However, for $m_X<2m_e$, finite temperature corrections become important when the $\ee$ interactions contribute significantly to the $X$ abundance, as discussed in section \ref{eXSec}.


\subsection{Boltzmann equations}
\label{boltzSec}

The distribution for a particle species $f$ evolves in accordance with the Liouville equation
\begin{equation}
\label{LiouvilleEqn}
\frac{\partial f}{\partial t}- H p \frac{\partial f}{\partial p} = \mathcal{C}[f], 
\end{equation}
for particle momentum $p$, Hubble expansion rate $H=\sqrt{8\pi \rho_{\text{tot}}/(3 M_{\rm Pl}^2)}$, and distribution dependent collision term $\mathcal{C}[f]$. The collision terms account for interactions affecting particle distributions, i.e., decays, annihilations, scattering, and their inverse processes.

With the approximations discussed in section~\ref{approxSec}, the Liouville equation in equation~(\ref{LiouvilleEqn}) can be reformulated in terms of the distribution's temperature and chemical potential,
\begin{subequations}
\label{boltzmannGeneralTmu}
\begin{align} 
\frac{dT_i}{dt}&=\mathbb{D}(T_i, \mu_i) \left[-3 H\left((\rho_i + P_i)\frac{\partial n_i}{\partial \mu_i} - n_i \frac{\partial \rho_i}{\partial \mu_i} \right)+\frac{\partial n_i}{\partial \mu_i}\frac{\delta \rho_i}{\delta t}-\frac{\partial \rho_i}{\partial \mu_i}\frac{\delta n_i}{\delta t} \right],\\ 
\frac{d\mu_i}{dt}&=-\mathbb{D}(T_i, \mu_i)
\left[-3 H\left((\rho_i + P_i)\frac{\partial n_i}{\partial T_i} - n_i \frac{\partial \rho_i}{\partial T_i} \right)+\frac{\partial n_i}{\partial T_i}\frac{\delta \rho_i}{\delta t}-\frac{\partial \rho_i}{\partial T_i}\frac{\delta n_i}{\delta t} \right],\\
\mathbb{D}(T, \mu)&=\left(\frac{\partial n}{\partial \mu}\frac{\partial \rho}{\partial T} - \frac{\partial n}{\partial T}\frac{\partial \rho}{\partial \mu} \right)^{-1}.
\end{align}
\end{subequations}
Here $n,\rho, P$ are the number, energy, and pressure densities for a particle $i$ with $d_i$ degrees of freedom obtained by integrating $f$ over $d_i \, d^3p/(2\pi)^3$, $d_i \, E \, d^3p/(2\pi)^3$, and $d_i \, \frac{p^2}{3E} \, d^3p/(2\pi)^3$ respectively. $\delta n/\delta t$ and $\delta \rho/\delta t$ are the number and energy transfer rates between particle species obtained by integrating $\mathcal{C}[f]$ over the same measures. The formulae for these thermodynamic quantities and their derivatives can be found in appendix~A.6 of \cite{EscuderoAbenza:2020cmq}. Explicitly, the seven Boltzmann equations parameterizing our system are given by
\begin{align} 
\label{sevenboltzmannGeneral}
\frac{dT_X}{dt}&= \mathbb{D}(T_X,\mu_X)\left[-3 H\left((\rho_X + P_X)\frac{\partial n_X}{\partial \mu_X} - n_X \frac{\partial \rho_X}{\partial \mu_X} \right)+\frac{\partial n_X}{\partial \mu_X}\frac{\delta \rho_X}{\delta t}-\frac{\partial \rho_X}{\partial \mu_X}\frac{\delta n_X}{\delta t} \right],\nonumber\\ 
\frac{dT_\nuR}{dt}&= \mathbb{D}(T_\nuR,\mu_\nuR)\left[-3 H\left((\rho_\nuR + P_\nuR)\frac{\partial n_\nuR}{\partial \mu_\nuR} - n_\nuR \frac{\partial \rho_\nuR}{\partial \mu_\nuR} \right)+\frac{\partial n_\nuR}{\partial \mu_\nuR}\frac{\delta \rho_\nuR}{\delta t}-\frac{\partial \rho_\nuR}{\partial \mu_\nuR}\frac{\delta n_\nuR}{\delta t} \right],\nonumber\\ 
\frac{dT_\nuL}{dt}&= \mathbb{D}(T_\nuL,\mu_\nuL)\left[-3 H\left((\rho_\nuL + P_\nuL)\frac{\partial n_\nuL}{\partial \mu_\nuL} - n_\nuL \frac{\partial \rho_\nuL}{\partial \mu_\nuL} \right)+\frac{\partial n_\nuL}{\partial \mu_\nuL}\frac{\delta \rho_\nuL}{\delta t}-\frac{\partial \rho_\nuL}{\partial \mu_\nuL}\frac{\delta n_\nuL}{\delta t} \right],\nonumber\\ 
\frac{dT_\gamma}{dt}&=\left(\frac{\partial \rho_\gamma}{\partial T_\gamma}+\frac{\partial \rho_e}{\partial T_\gamma}\right)^{-1}\left[ 4H\rho_\gamma +3H(\rho_e+P_e)+\frac{\delta\rho_e}{\delta t}\right],\nonumber\\
\frac{d\mu_X}{dt}&= -\mathbb{D}(T_X,\mu_X)\left[-3 H\left((\rho_X + P_X)\frac{\partial n_X}{\partial T_X} - n_X \frac{\partial \rho_X}{\partial T_X} \right)+\frac{\partial n_X}{\partial T_X}\frac{\delta \rho_X}{\delta t}-\frac{\partial \rho_X}{\partial T_X}\frac{\delta n_X}{\delta t} \right],\nonumber\\
\frac{d\mu_\nuR}{dt}&= -\mathbb{D}(T_\nuR,\mu_\nuR)\left[-3 H\left((\rho_\nuR + P_\nuR)\frac{\partial n_\nuR}{\partial T_\nuR} - n_\nuR \frac{\partial \rho_\nuR}{\partial T_\nuR} \right)+\frac{\partial n_\nuR}{\partial T_\nuR}\frac{\delta \rho_\nuR}{\delta t}-\frac{\partial \rho_\nuR}{\partial T_\nuR}\frac{\delta n_\nuR}{\delta t} \right],\nonumber\\
\frac{d\mu_\nuL}{dt}&= -\mathbb{D}(T_\nuL,\mu_\nuL)\left[-3 H\left((\rho_\nuL + P_\nuL)\frac{\partial n_\nuL}{\partial T_\nuL} - n_\nuL \frac{\partial \rho_\nuL}{\partial T_\nuL} \right)+\frac{\partial n_\nuL}{\partial T_\nuL}\frac{\delta \rho_\nuL}{\delta t}-\frac{\partial \rho_\nuL}{\partial T_\nuL}\frac{\delta n_\nuL}{\delta t} \right].
\end{align}
The $\nu_R$ equations are included only in the Dirac case. The transfer rates $\delta n_i/\delta t$ and $\delta \rho_i/\delta t$ will be discussed in the next section.


\subsection{Collision terms}
\label{collisionSec}

\begin{figure}
\begin{center}
\includegraphics[width=0.95\textwidth]{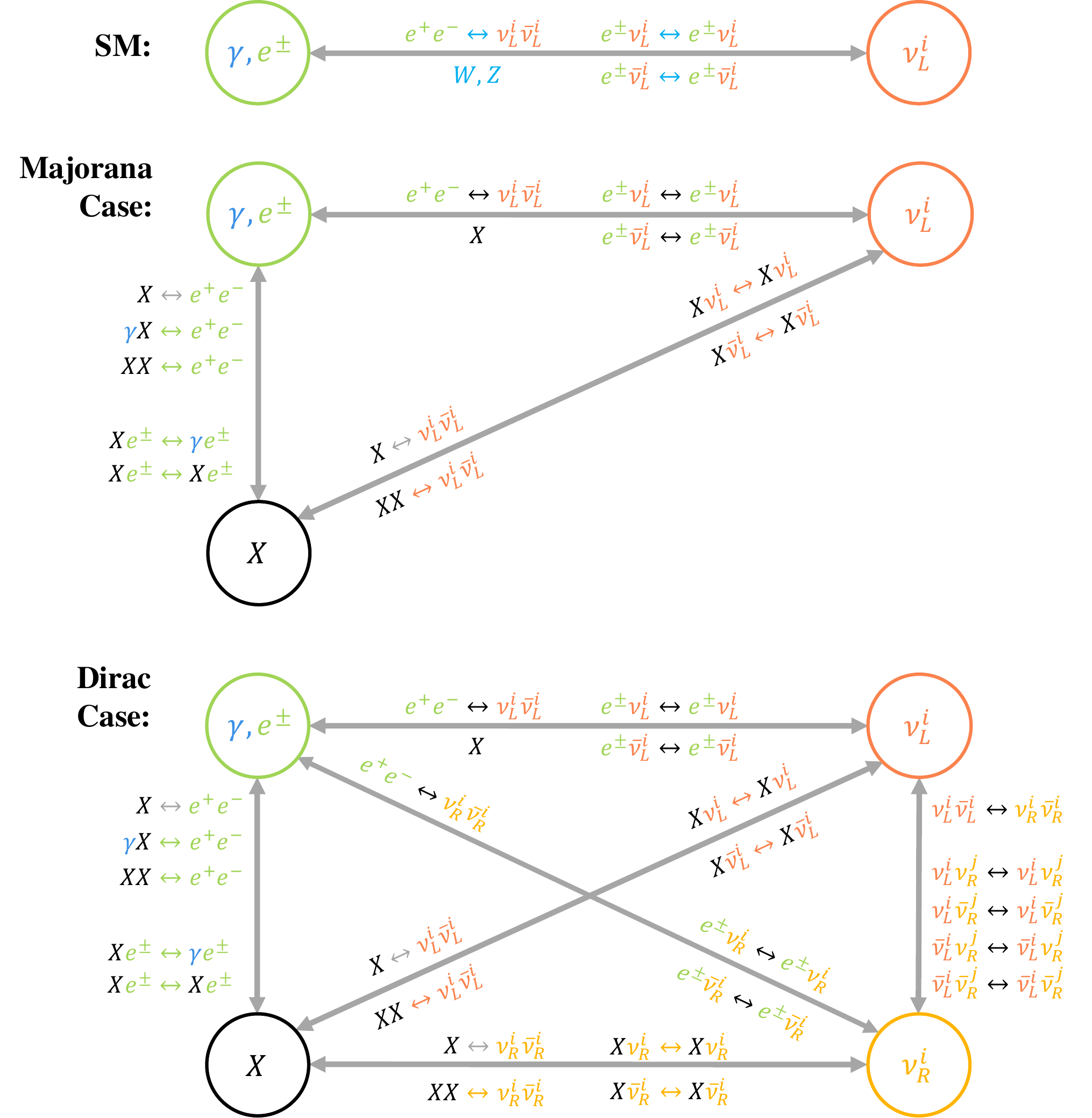}
\end{center}
\caption{Summary of the processes considered in deriving the collision terms that relate different thermodynamic sectors in our model. All of the $1 \leftrightarrow 2$ processes are renormalizable contact interactions, while the $2 \leftrightarrow 2$ processes include all $s$,$t$, and $u$-channel exchange diagrams that preserve lepton number, electric charge, and flavor. 
  Arrow color implies the mediator (or absence thereof):
  $X$ (black); $W$, $Z$ (light blue); $\nu_L^i$ (red); $\nu_R^i$ (yellow);
  and $1 \leftrightarrow 2$ contact interaction (grey).
  }
\label{particleProcessesFig}
\end{figure}

\subsubsection{Electron-neutrino interactions}
\label{eNuSec}

In the SM, the relevant interactions between electrons and left-handed neutrinos are the weak interactions $e^{+}e^{-}\lra\bar{\nu}_\alpha\nu_\alpha, \; e^{\pm}\nu_\alpha \lra e^{\pm}\nu_\alpha$, and $e^{\pm}\bar{\nu}_\alpha \lra e^{\pm}\bar{\nu}_\alpha$. The transfer rates for these processes are given by \cite{EscuderoAbenza:2020cmq}
\begin{subequations}
\label{nuSMTransferRates}
 \begin{align}
\frac{\delta \rho^{\smT}_{e-\nu}}{\delta t}&=2\frac{G_F^2}{3\pi^5}\left[C_V^2+C_A^2+2\left(\tilde{C}_V^2+\tilde{C}_A^2\right)\right]F_\rho(T_\gamma,T_\nu,\mu_\nu)\\
\frac{\delta n^{\smT}_{e-\nu}}{\delta t} &=16\frac{G_F^2}{3\pi^5}\left[C_V^2+C_A^2+2\left(\tilde{C}_V^2+\tilde{C}_A^2\right)\right]F_n(T_\gamma,T_\nu,\mu_\nu) \, .
\end{align}
\end{subequations}
Here we defined 
\begin{subequations}
\label{FtempDiffDefinition}
\begin{align}
  F_\rho(T_\gamma,T_\nu,\mu_\nu)&=32\,(T_\gamma^9-T_\nu^9 e^{\frac{2\mu_\nu}{T_\nu}})+56\, T_\gamma^4T_\nu^4 e^{\frac{\mu_\nu}{T_\nu}} \left(T_\gamma-T_\nu\right), \\
F_n(T_\gamma,T_\nu,\mu_\nu)&=T_\gamma^8-T_\nu^8e^{\frac{2\mu_\nu}{T_\nu}}.
\end{align}
\end{subequations}
with $C_V=g^{\nu_e}_L+g^{\nu_e}_R=\frac{1}{2}+2s_W^2$, $C_A=g^{\nu_e}_L-g^{\nu_e}_R=\frac{1}{2}$, $\tilde{C}_V=g^{\nu_{\mu,\tau}}_L+g^{\nu_{\mu,\tau}}_R=-\frac{1}{2}+2s_W^2$, and $\tilde{C}_A=g^{\nu_{\mu,\tau}}_L-g^{\nu_{\mu,\tau}}_R=-\frac{1}{2}$ being the weak vector and axial couplings, $s_W^2 \simeq 0.223$ is the weak mixing angle, and $G_F \simeq 1.16 \times 10^{-11} \,\mev^{-2}$ is the Fermi constant. 

These interactions mediated by weak boson exchange
can also be mediated by $X$-exchange.  Since the weak interactions
drop out-of-equilibrium near $T\sim 1$~MeV, we are interested in
the strength of $X$-exchange interactions when $X$ can also be
integrated out.  For 
$m_X \gtrsim 10$~MeV, we can write the effective interaction Lagrangian 
\begin{equation} 
\label{integratedOutXLagrangina}
\mathcal{L}_{\text{int}} \;=\; -\frac{g_X^2}{m_X^2}\left[\left(\bar{\nu}_L \gamma^\alpha \nu_L + \bar{\nu}_R\gamma^\alpha \nu_R\right) \right](\bar{e}\gamma_\alpha e) \, .
\end{equation}
The transfer rates now have the same form as equation~(\ref{nuSMTransferRates}) that only differs by the overall coefficient.  Given that $\tilde{C}_V$ and $\tilde{C}_A$ correspond to the coefficients for just $Z$-exchange in the SM, we can obtain the transfer rate for $X$-exchange by setting $C_V=C_A=0$ and $2 (\tilde{C}_V^2 + \tilde{C}_A^2) \ra 3 q_X^2$, where $X$-exchange proceeds through a purely vector interaction with charge $q_X^2 = 1$ for each lepton flavor, and $G_F\rightarrow \frac{g_X^2}{2\sqrt{2} m_X^2}$.
The full $e-\nu$ transfer rates are
 \begin{subequations}
\label{nuSMPlusXTransferRates}
 \begin{align}
\frac{\delta \rho_{e-\nu}}{\delta t} &= \frac{1}{3\pi^5} \left[ G_F^2
\left(3-4\,s_W^2+24\,s_W^4\right)
+\frac{3}{4}\frac{g_X^4}{m_X^4}
\right] F_\rho(T_\gamma,T_\nu,\mu_\nu)  \\
\frac{\delta n_{e-\nu}}{\delta t}&= \frac{8}{3\pi^5} \left[ G_F^2
\left(3-4\,s_W^2+24\,s_W^4\right)
+\frac{3}{4} \frac{g_X^4}{m_X^4}
\right] F_n(T_\gamma,T_\nu,\mu_\nu) \, .
\end{align}
\end{subequations}
The $X$-exchange interaction can overpower the SM weak process when $g_X^2/m_X^2\sim \mathcal{O}(G_F) \sim 10^{-11} \,\mev^{-2}$. In this case, electron-neutrino (proton-neutron) decoupling is delayed to temperatures lower than that predicted in standard cosmology.  As an extreme example, if the neutrino decoupling temperature was pushed below the electron-positron annihilation regime, then  $T_\nu=T_\gamma$ today [as opposed to $T_\nu\sim\left(4/11\right)^{1/3}T_\gamma$], which
would result in $\neff \simeq 3 (11/4)^{4/3} \simeq 11.6$


\subsubsection{Decays and inverse decays}
\label{decaySec}

When kinematically accessible, the dominant BSM processes are $X\lra \ee$,  $X\lra \nunuL$, and $X\lra \nunuR$.
The partial decay widths of $X$ to electron-positron or neutrino pair (left- or right-handed, one generation) is given by
\begin{equation}
\label{partialDecayWidthToee}
\Gamma_{X\rightarrow \ee} =\frac{g_X^2 }{12\pi}m_X\left(1+2\frac{m_e^2}{m_X^2}\right) \sqrt{1-4\frac{m_e^2}{m_X^2}} \quad \quad \quad \quad  \Gamma_{X\rightarrow \nunui} =\frac{g_X^2 }{24\pi}m_X.
\end{equation}
In Eq.~\ref{partialDecayWidthToee}, we use the vacuum value for the electron mass since we have verified that the finite temperature corrections shift our results only by a very small (at most $1$-$2$\%) amount (see Sec.~\ref{approxSec} for details on our approximations).
The collision term for the decay and inverse decay $X\lra \aab$ is
\begin{equation} 
\label{decayCollisionTerm}
\mathcal{C}_{X\rightarrow \aab}(p_X)=-\Gamma_{X\rightarrow \aab}\frac{m_X}{m_*}\frac{m_X}{E_X p_X} \int^{E_{+}}_{E_{-}} dE_a \left[f_X(1-f_a)(1-f_a)-f_af_a(1+f_X)\right]
\end{equation}
with $E_X^2 = p_X^2 + m_X^2$, $m_*^2=m_X^2-4m_a^2$ and the bounds $E_\pm=\frac{1}{2}[E_X\pm (p_X m_*)/m_X ]$. For example in $X\lra \nu \bar{\nu}$, the first term in equation~(\ref{decayCollisionTerm})
is explicitly
\begin{equation}  
\label{decayDistributionFunctions}
f_X(1-f_a)(1-f_a)\longrightarrow f_X(E_X,T_X,\mu_X)\left[1-f_\nu(E_\nu,T_\nu,\mu_\nu)\right]\left[1-f_\nu(E_X-E_\nu,T_\nu,\mu_\nu)\right].
\end{equation}
The integral in equation~(\ref{decayCollisionTerm}), can be solved analytically
taking a Bose-Einstein distribution for $X$ and a
Fermi-Dirac distribution for neutrinos and electrons, 
giving
\begin{equation}  
\label{integratedDecayCollisionTerm}
\mathcal{C}_{X\rightarrow \aab}=-\Gamma_{X}\frac{m_X}{m_*}\frac{m_X T_a}{E_X p_X}
\frac{e^{\frac{E_X}{T_a}+\frac{\mu_X}{T_X}}-e^{\frac{E_X}{T_X}+\frac{2\mu_a}{T_a}}  }{\left(e^{\frac{E_X}{T_a}}-e^{\frac{2\mu_a}{T_a}} \right)\left(e^{\frac{E_X}{T_X}}-e^{\frac{\mu_X}{T_X}} \right)}\log\left[
\frac{\left(e^{\frac{E_X}{T_a}}+e^{\frac{E_{-}+\mu_a}{T_a}} \right)
\left(e^{\frac{E_{+}}{T_a}}+e^{\frac{\mu_a}{T_a}} \right)}{
\left(e^{\frac{E_X}{T_a}}+e^{\frac{E_{+}+\mu_a}{T_a}} \right)
\left(e^{\frac{E_{-}}{T_a}}+e^{\frac{\mu_a}{T_a}} \right)}
\right].
\end{equation}
Finally, the decay and inverse decay transfer terms are 
\begin{subequations}
\label{nETransferRates}
\begin{align} 
\frac{\delta\rho_{X\lra \aab}}{\delta t}&= 
\frac{d_X}{(2\pi)^3} \int d^3p_X E_X\mathcal{C}_{X\rightarrow \aab}(p_X)
\\
\label{numDenDecayTransferRate}
\frac{\delta n_{X\lra \aab}}{\delta t}&=
\frac{d_X}{(2\pi)^3} \int d^3p_X\; \mathcal{C}_{X\rightarrow \aab}(p_X)
\end{align}
\end{subequations}
where $d_X=3$ is the number of spin degrees of freedom of $X$.


\subsubsection{Electron-X interactions}
\label{eXSec}

When $m_X<2m_e$, the $X\lra \ee$ process is kinematically forbidden, and so the available $e$-$X$ interactions are $\gamma X\lra\ee$ and $e^{\pm} X\lra e^{\pm}\gamma$.  These transfer rates can be derived following appendix~A.7 of \cite{EscuderoAbenza:2020cmq} giving,
\begin{subequations}
\label{2To2CollisionTerm}
\begin{align} 
\frac{\delta\rho_{2\lra2}}{\delta t}&=-\frac{d_1d_2}{64\pi^4} \int^{\infty}_{s_{\text{min}}} ds\;  s \,\sigma_{2\lra2}(s)\left[s\, T_\gamma\, K_2\left(\frac{\sqrt{s}}{T_\gamma}\right)- \mathcal{A}(s,T_\gamma,T_X,\mu_X) \right],\\
\frac{\delta n_{2\lra2}}{\delta t}&=-\frac{d_1d_2}{64\pi^4} \int^{\infty}_{s_{\text{min}}} ds\;  s \,\sigma_{2\lra2}(s)\left[2\sqrt{s} \, T_\gamma\, K_1\left(\frac{\sqrt{s}}{T_\gamma}\right)- \mathcal{B}(s,T_\gamma,T_X,\mu_X) \right], 
\end{align}
\end{subequations}
where $d_1, d_2$ are the number of spin degrees of freedom of incoming particles $1$ and $2$, and $\sigma_{2\lra2}$ is the cross section of the aforementioned interactions given in appendix~C of \cite{Redondo:2008ec}, and 
\begin{subequations}
\label{2To2AB}
\begin{align} 
\mathcal{A}(s)&=e^{\frac{\mu_X}{T_X}}\int dE_{+}dE_{-} \frac{E_{+}+E_{-}}{2}\exp\left[{-\frac{E_{+}-E_{-}}{2T_\gamma}}\right]\exp\left[{-\frac{E_{+}+E_{-}}{2T_X}}\right]\\
\mathcal{B}(s)&=e^{\frac{\mu_X}{T_X}}\int dE_{+}dE_{-} \exp\left[{-\frac{E_{+}-E_{-}}{2T_\gamma}}\right]\exp\left[{-\frac{E_{+}+E_{-}}{2T_X}}\right]
\end{align} 
\end{subequations}
with $|E_{-}-E_{+}(m_2^2-m_1^2)/s|\leq 2 p_{12}\sqrt{(E_{+}^2-s)/s}$,  $p_{12}=[s-(m_1+m_2)^2]^{1/2}[s-(m_1-m_2)^2]^{1/2}/(2\sqrt{s})$, and $E_{+}\geq \sqrt{s}$. Part of Eqs.~(\ref{2To2CollisionTerm}),(\ref{2To2AB}) can be evaluated analytically, while the remainder must be done numerically.  To further simply our evaluation of the phase space integrals in equation~(\ref{2To2AB}), we have taken $m_X = 0$, thereby slightly overestimating the integration region.  In practice, this overestimate is only possibly relevant when $m_X \sim 2 m_e$, and in this region, the $2 \ra 2$ processes do not set
the strongest bounds.

Finite temperature effects make significant corrections to these interactions, as we have explained in detail using analytic arguments presented in appendix \ref{app:finiteTemperatureEffects}.
In our Boltzmann evolution, we include these effects by replacing the vacuum coupling $g_X$  in $\sigma_{2\lra2}$ by the in-medium coupling 
\begin{equation}  
\label{inMediumCouplingText}
g_X^2 \longrightarrow g_{X,m}^2 = \frac{g_X^2 m_X^4}{[m_X^2 -m_\gamma(T)^2]^2 + [\text{Im} \, \Pi (T) ]^2},
\end{equation}
where $m_\gamma$ is the in-medium effective photon mass and $\text{Im} \, \Pi $ is the imaginary part of the photon polarization tensor.  This demonstrates that there will be a \emph{resonant enhancement} that occurs when
$m_X = m_\gamma(T)$ at a specific temperature $T_r$ for a given $m_X$.
The in medium photon mass $m_\gamma(T)$ is calculated using the formalism in \cite{Braaten:1993jw} for the real part of the photon polarization tensor. It should be noted that $m_\gamma(T)$ is also dependent on the photon energy, however, this weak dependence does not alter any dynamics and can be neglected. Our numerical evaluation of $m_\gamma(T)$ matches Eq.~\ref{effectiveMassesOfPhoton} in the limits shown, $T \gg m_e$ and $T \ll m_e$. Next, we take $\text{Im} \, \Pi (T) =- \omega  \Gamma$ as in Eq. \ref{ImPolarizationResponse} \cite{Weldon:1983jn}. We consider only Compton scattering and so $\Gamma = 8\pi\alpha^2/(3m_e^2)n_e$ (for $T\ll m_e$), $\Gamma \sim \alpha^2 T^2/(\pi \omega)\log(4T\omega/m_e^2)$ (for $T\gg m_e$) \cite{Redondo:2008ec}, and then take a linear interpolation between these two limiting expressions. Note that the kink in the $\dneff = 0.4,0.5$ contours in figure~\ref{dneffMFig} and $\dneff = 0.2$ and larger contours in figure~\ref{dneffDFig} around $m_X \sim 50 \;\kev$ are a result of this linear interpolation. The contours on either side of the kink are accurate.


\subsubsection{Suppressed interactions}
\label{otherInterSec}

In Table~\ref{particleProcessesFig}, we have shown several other
interactions between $X$ and the SM as well as interactions mediated
by (virtual) $X$-exchange.  We now briefly summarize the suppressions
that these interactions have, and why we can neglect them in our
evaluation of the Boltzmann evolution.

The interaction $\ee \lra XX$ is suppressed by a factor of $g_X^2/e^2$ compared to $\ee \lra \gamma X$.  For the neutrino-mediated processes $\nunuL \lra XX$ and $\nunuR  \lra XX$, the interaction rate $n_\nu \langle\sigma v\rangle_{\nunui \lra XX}$ is suppressed by an additional power of $g_X^2$ as  compared to equivalent decay channels $X\lra \nunui$. Similarly, $X \epm \lra X \epm, \;\;X\nu_L \lra X\nu_L,  \;\; X\nu_R \lra X\nu_R$ are also suppressed by $g_X^2$ compared to the (inverse) decay processes, but in addition, also do not contribute to the change in number density of individual particle species. These interactions can be safely neglected.

In the Dirac neutrino case, there are additional processes involving right-handed neutrinos that arise, similar to the ones discussed in section~\ref{eNuSec}. These include $e^{+}e^{-}\lra\bar{\nu}_R\nu_R, \; e^{\pm}\nu_R \lra e^{\pm}\nu_R$,  $e^{\pm}\bar{\nu}_R \lra e^{\pm}\bar{\nu}_R$ in the $e$-$\nu_R$ system and $\nunuL\lra\nunuR,\;\; \nu_L \bar{\nu}_R \lra \nu_L \bar{\nu}_R$, and $\nu_L \nu_R \lra \nu_L \nu_R$ in the $\nu_L$-$\nu_R$ system.
We did not include $X$-exchange to right-handed neutrinos for the same reasons as above, namely, they are further suppressed by an additional power of $g_X^2$.


\subsection{Summary of transfer rates}
\label{transferSec}

Following the discussion of the collision terms, section~\ref{collisionSec},
we now explicitly provide the transfer rates
$\delta \rho_i/\delta t$ that we use in equation~(\ref{sevenboltzmannGeneral}):
\begin{subequations}
\label{allTransferRatesEq}
\begin{align} 
\frac{\delta\rho_{X}}{\delta t}&=\frac{\delta\rho_{X\lra \ee}}{\delta t}+\frac{\delta\rho_{2\lra2}}{\delta t}+N_\nu \left[\frac{\delta\rho_{X\lra \nunuL}}{\delta t}+ \left\{\frac{\delta\rho_{X\lra \nunuR}}{\delta t} \right\}\right]
,\\
\frac{\delta\rho_{\nu_R}}{\delta t}&=\left\{-\frac{1}{2} \frac{\delta\rho_{X\lra \nunuR}}{\delta t} \right\}
\label{nuRTransferEq}
,\\
\frac{\delta\rho_{\nu_L}}{\delta t}&=\frac{1}{2} \left[\frac{\delta \rho_{e-\nu}}{\delta t}-\frac{\delta\rho_{X\lra \nunuL}}{\delta t}\right]
\label{nuLTransferEq}
,\\
\frac{\delta\rho_{\gamma}}{\delta t}&=-N_\nu \frac{\delta \rho_{e-\nu}}{\delta t}-\frac{\delta\rho_{X\lra \ee}}{\delta t}-\frac{\delta\rho_{2\lra2}}{\delta t} \, .
\end{align}
\end{subequations}
Note that the neutrino Boltzmann equations in equation~(\ref{sevenboltzmannGeneral}) evolve a single neutrino species (rather neutrino and anti-neutrino) which accounts for the factor of $1/2$ in Eqs.~(\ref{nuRTransferEq})-(\ref{nuLTransferEq}). 
The $\delta n_i/\delta t$ rates can be straightforwardly deduced from Eqs.~(\ref{allTransferRatesEq}).


\section[Numerical evolution of the Boltzmann equations and \texorpdfstring{$\neff$}{Neff}]{Numerical evolution of the Boltzmann equations and $N_{\rm eff}$}
\label{sec:numericalevolutionneff}

We now discuss the details of solving the Boltzmann equations
in equation~(\ref{sevenboltzmannGeneral}) in order to determine the
thermodynamic evolution of the universe with $X$ interactions.
The final result from this section, calculating the contours of
$\Delta N_{\rm eff}$ in the $(m_X, g_X)$ parameter space,
is the central result of this paper.

\subsection{Initial conditions}

We start with the initial condition
$T_0\equiv T_\gamma=T_{\nu_L}= \, {\rm Max}[10 \; \mev, \; 10 \, m_X]$.
and $\mu_{\nu_L} = - 10^{-5} \,T_0$.  This ensures that all the dynamics before neutrino decoupling are captured.  Furthermore, we start $T_X=T_{\nu_R}=10^{-2}\,T_0$ and $\mu_X=\mu_{\nu_R}=-10^{-5}\,T_0$, so that  both $X$ and $\nu_R$ begin with a negligible abundance compared to the SM bath, $\rho_X/\rho_\nu \sim 10^{-8}$. The specific (small) initial values of $T_X, \mu_X$ and their relation to $T_{\nu_R}, \mu_{\nu_R}$ do not impact the evolution, since the Boltzmann equations quickly evolve these quantities to their correct values. 

The evolution of particle distributions is often shown as $\rho_i/T_\gamma^4$,  using $T_\gamma$ to show dimensionless ratios of energy densities of difference species.  This is useful because in evolving over several orders of magnitude in temperature, a relativistic species energy density spans \emph{many} orders of magnitude.  Moreover, in cases where new physics does not affect photons,
e.g.~\cite{Escudero:2019gzq,EscuderoAbenza:2020cmq},
the evolution of $T_\gamma$ can be used to provide a convenient measure of time that is independent of the new physics dynamics.

In our case, $X$ interactions affect the evolution of photons, electrons, and neutrinos distributions.  This means that in trying to compare the energy densities and number densities of different species with different choices of parameters ($m_X, g_X$), there can be differential effects on the evolution of $T_\gamma$, and thus both the numerator $\rho_i$ and the denominator $T_\gamma^4$. 
We therefore introduce a ``reference'' temperature, $T_\xi$, obtained by solving the Boltzmann equation $dT_\xi/dt = -H T_\xi$ with
$T_\xi(t_0)=T_0$.\footnote{One way to think about this is to imagine $\xi$ is a (fictitious) massless particle with a temperature that is initially set to the temperature of the SM but is actually decoupled from all other particles.  Taking $\xi$ to be massless with an infinitesimal small number of degrees of freedom means $\xi$ always evolves as radiation and does not affect Hubble expansion rate ($\rho_\xi \sim 0$).}
In the following, we show energy and number densities scaled by $T_\xi^4$ or $T_\xi^3$ respectively, and this allows us to much more easily compare different choices of $(m_X, g_X)$ with each other.


\subsection{Evolution of Boltzmann equations}
\label{sec:evolutionboltzmann}

In this section, we discuss the solution to the Boltzmann equations in equation~(\ref{sevenboltzmannGeneral}) at representative points in the parameter space. To do so, we examine the evolution of density and effective interaction rate for the particles in our system. The effective interaction rate\footnote{In \cite{EscuderoAbenza:2020cmq}, this quantity is instead defined as $(\delta n_{X\rightarrow\nunui}\delta t)/(H n_\nu)$ for the one-way interaction. This is done to estimate the interaction strength at which $X$ thermalizes with neutrinos, $\langle \Gamma_i\rangle \sim H$, without solving the Boltzmann equations.  We replace $n_\nu$ by $T_\xi^3$ as we are also interested in electron interactions and $n_e \rightarrow 0$ after electron-positron annihilation. Secondly, we consider the forward and backward interaction in the transfer rates instead of just the forward rate [first term in square bracket of equation~(\ref{numDenDecayTransferRate})].} is defined as 
\begin{equation} 
\label{effectiveInteractionStrength}
\frac{\langle\Gamma_i \rangle}{H}\equiv \frac{1}{HT_\xi^3}\frac{\delta n_i}{\delta t}
\end{equation}
where $\delta n_i/\delta t$ is the sum of all number density transfer rates (section~\ref{transferSec}) for interactions between two species $i=X\lra e$ or $X\lra \nu$.  This is the rate to which the forward or backward interaction dominate the other as compared to Hubble expansion rate. For example, a positive $\langle\Gamma_{X\lra \nu} \rangle/H$ means the $\nunui \rightarrow X$ dominates the backward rate and the neutrinos number density is decreasing in favor of $X$. Similarly, zero means forward/backward rates are equal and a zero plateau implies that $X$ and neutrinos are thermalized while $X$ is relativistic.

\begin{figure}
\begin{center}
\includegraphics[width=1.0\textwidth]{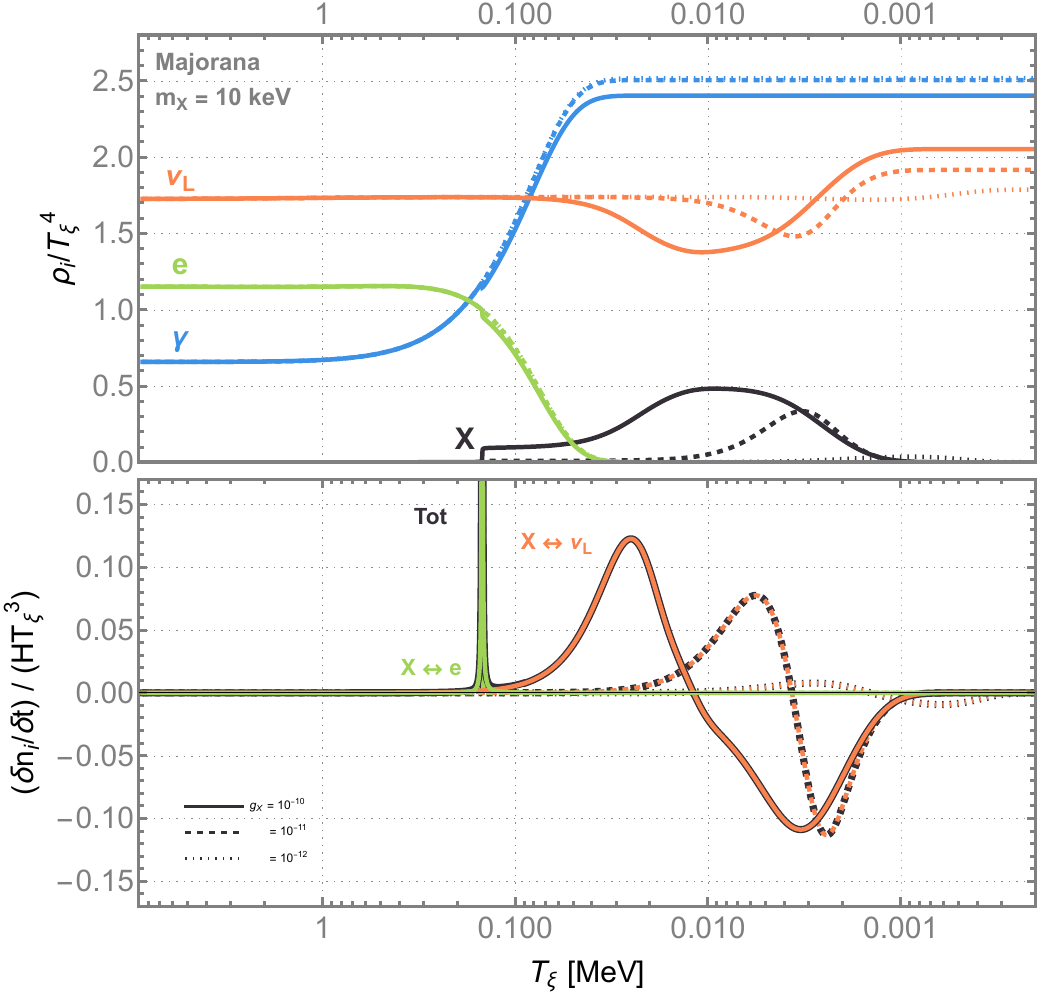}
\caption{ \textit{Top:} The scaled density and \textit{bottom:}
  and effective interaction rate for $m_X=10\,\kev$ for couplings
  $g_X=10^{-10}$ (solid, $\dneff=0.72$),
  $g_X=10^{-11}$ (dashed, $\dneff=0.33$), and
  $g_X=10^{-12}$ (dotted, $\dneff=0.08$).} 
\label{m1EvolutionFig}
\end{center}
\end{figure}

We first consider an example of an $X$ boson that is light, $m_X=10$~keV,
that is representative of the light regime $m_X \ll 2 m_e$.  
This is among the simpler cases since the $X\lra \ee$ process is kinematically
forbidden and for sufficiently low coupling the only relevant interaction is
$X \lra \nunuL$.  In figure~\ref{m1EvolutionFig},
we show the scaled density and effective interaction rates for several values of $g_X = 10^{-10}$ to $10^{-12}$.  At the smaller end of the couplings, $g_X=10^{-11}$ and $g_X=10^{-12}$, neutrinos populate $X$ without fully thermalizing (at $g_X=10^{-11}$, $ X \lra \nunuL$ is almost thermalized).  $X$ then proceeds to evolve as matter $\rho_X \propto a^{-3}$ before dumping its entropy back into neutrinos. The nonrelativistic $X$'s decay into neutrinos that are now more energetic than the primordial neutrino plasma that evolves as radiation $\rho_\nu \propto a^{-4}$. Therefore, the resulting neutrino energy density is larger than it would have been in the SM\@.  This implies a positive $\neff$ is generated, as shown in the figure caption.

Also, in figure~\ref{m1EvolutionFig}, for $g_X=10^{-10}$, the most obvious feature is the resonance near $T_\xi \sim 0.15$~MeV resulting from the finite temperature corrections to $\gamma X\lra\ee$ and $e^{\pm} X\lra e^{\pm}\gamma$.  In fact, the resonance is also present at the same temperature for the smaller values $g_X=10^{-11},10^{-12}$, but it is only barely visible in the plot given the smaller coupling.  This results in a finite contribution to $X$ shown by the step-up its energy density, and then $\nunuiL \rightarrow X$ also becomes efficient at a later time, resulting in the bump in production of $X$ near $T_\xi \sim 0.1$~MeV\@.  
%
%
%
%
Note that the $X \lra e$ effective interaction rate is always positive since $m_e>m_X$ and electron-positron annihilation occurs before $X$ decay.  This means that the forward reactions ($\gamma X \ra \ee$ and
$e^{\pm} X \ra e^{\pm}\gamma$) never dominates over the backward reactions.  So while electrons transfer their energy to $X$, the energy is never returned (unlike the case with neutrinos). This affects the evolution two-fold: (i) electron-positron annihilation dumps less entropy into photon plasma, decreasing $\rho_\gamma$; (ii) entropy stolen from electrons/positrons is later dumped into neutrinos, further increasing $\rho_\nu$. Both effects result in a larger value of $\neff$.

\begin{figure}
\begin{center}
\includegraphics[width=1.0\textwidth]{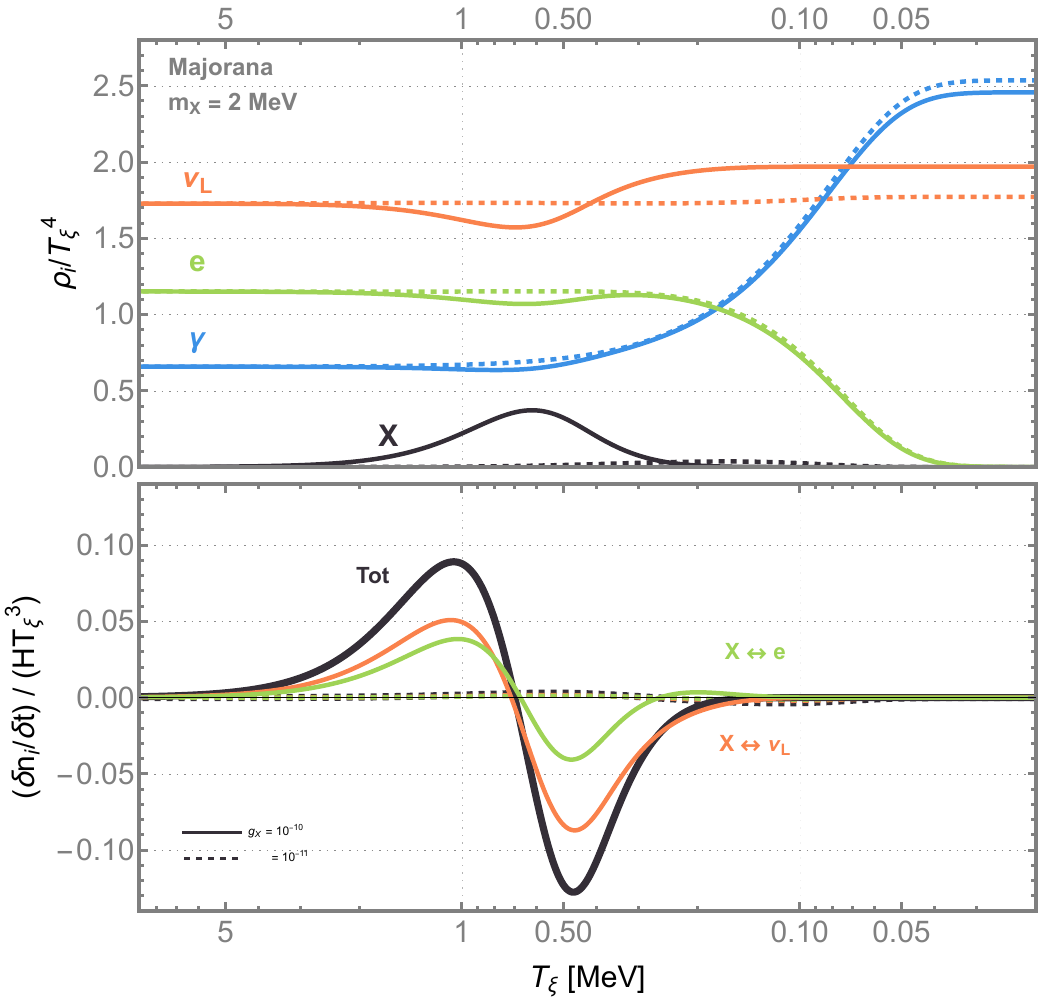}
\caption{\textit{Top:} The scaled density and \textit{bottom:}
  and effective interaction rate for $m_X=2\,\mev$ for couplings
  $g_X=10^{-10}$ (solid, $\dneff=0.49$) and
  $g_X=10^{-11}$ (dashed, $\dneff=0.03$).}
\label{p3EvolutionFig}
\end{center}
\end{figure}
\begin{figure}
\begin{center}
\includegraphics[width=1.0\textwidth]{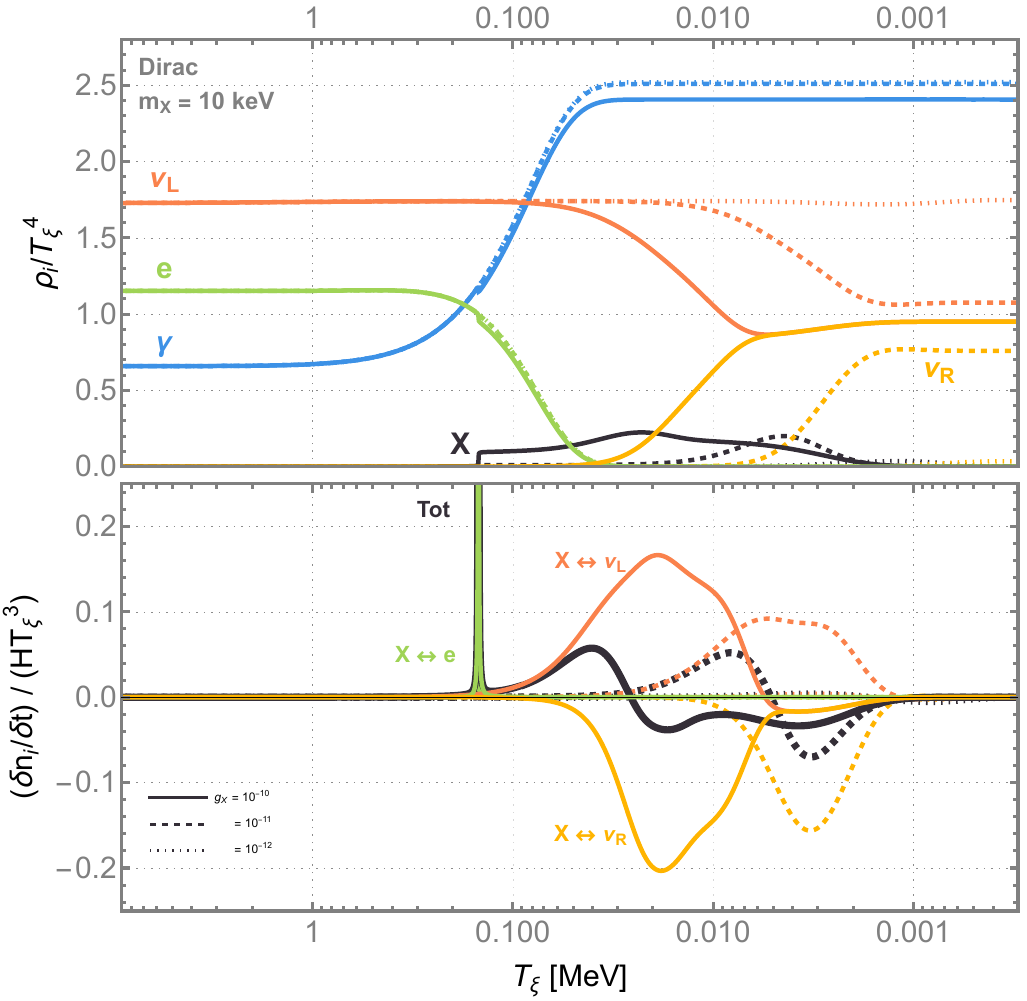}
\caption{\textit{Top:} The scaled density and \textit{bottom:} and effective
  interaction rate for $m_X=10\,\kev$ with Dirac neutrinos for couplings
  $g_X=10^{-10}$ (solid, $\dneff=0.43$),
  $g_X=10^{-11}$ (dashed, $\dneff=0.18$), and
  in the upper figure only, $g_X=10^{-12}$ (dotted, $\dneff=0.07$).}
\label{p3RIGHTEvolutionFig}
\end{center}
\end{figure}

In figure~\ref{p3EvolutionFig}, 
we show the same plot for $m_X=2$~MeV\@.  In this case, the $X\lra \ee$ process is active and the population of $X$ particles is depleted (the decays of $X$s) before electron-positron annihilation has completed. 
For $g_X=10^{-10}$, electrons and neutrinos populate a distribution of $X$ until $T_\xi \simeq 0.7$~MeV, then $X$ decay takes over and dumps its entropy back into both electrons and neutrinos. However, for this mass, both $e^\pm$ and $X$ become non-relativistic and evolve as matter around roughly the same time. This means that the back and forth exchange in energy does not comparatively increase the energy density of electrons as it does for neutrinos. Another effect shown in figure~\ref{p3EvolutionFig} (lower) is the evolution of the effective $X \lra e$ interaction rate from positive to negative, and then back to positive. The first positive bump is the forward reaction $\ee \ra X$,
followed by $X$ going nonrelativistic, with $X$ decaying into $\ee$ and $\nunuL$, and then finally $\ee \ra X$ occurs on the Boltzmann tail of the electron distribution, and this small regenerated population of $X$ decays mostly back into neutrinos.

In figure~\ref{p3RIGHTEvolutionFig},
we show the evolution for $m_X=10$~keV with Dirac neutrinos. For low couplings, $\nunuiL \rightarrow X$ populates $X$ and $X\rightarrow \nunuiR$ populates right-handed neutrinos driving $X -\nuL - \nuR$ system into thermal equilibrium. The energy density of primordial left-handed neutrinos is now distributed among $X$, $\nuL$, and $\nuR$.
Hence, the maximum energy density that $X$ can achieve is smaller than in the Majorana neutrino case. Increasing the coupling from $g_X=10^{-12}$ to $g_X=10^{-10}$ shows the trend towards thermalization of $\nuL$ with $\nuR$ as the interaction rate becomes efficient. For $g_X=10^{-10}$, the resonance in the $X \lra e$ system is, again, the most prominent feature in the plot. The dynamics of resonance are identical to those in figure~\ref{m1EvolutionFig}. Furthermore, for $g_X=10^{-10}$, the initial distribution of left-handed neutrinos has been converted to an equal distribution of left- and right-handed neutrinos. If this were to happen instantaneously, there would be no effect on $\neff \propto (\rho_\nuL+\rho_\nuR)/\rho_\gamma$. However, since this thermalization process generates an abundance of $X$, which then becomes nonrelativistic, evolves as matter, and then decays, there is an increase in $N_{\rm eff}$ just as occurred  in the Majorana case in figure~\ref{m1EvolutionFig}.


\subsection[Numerical results for \texorpdfstring{$\neff$}{Neff}]{Numerical results for $N_{\rm eff}$}
\label{cnstrntSec}

\begin{figure}[t]
\begin{center}
\includegraphics[width=1.0\linewidth]{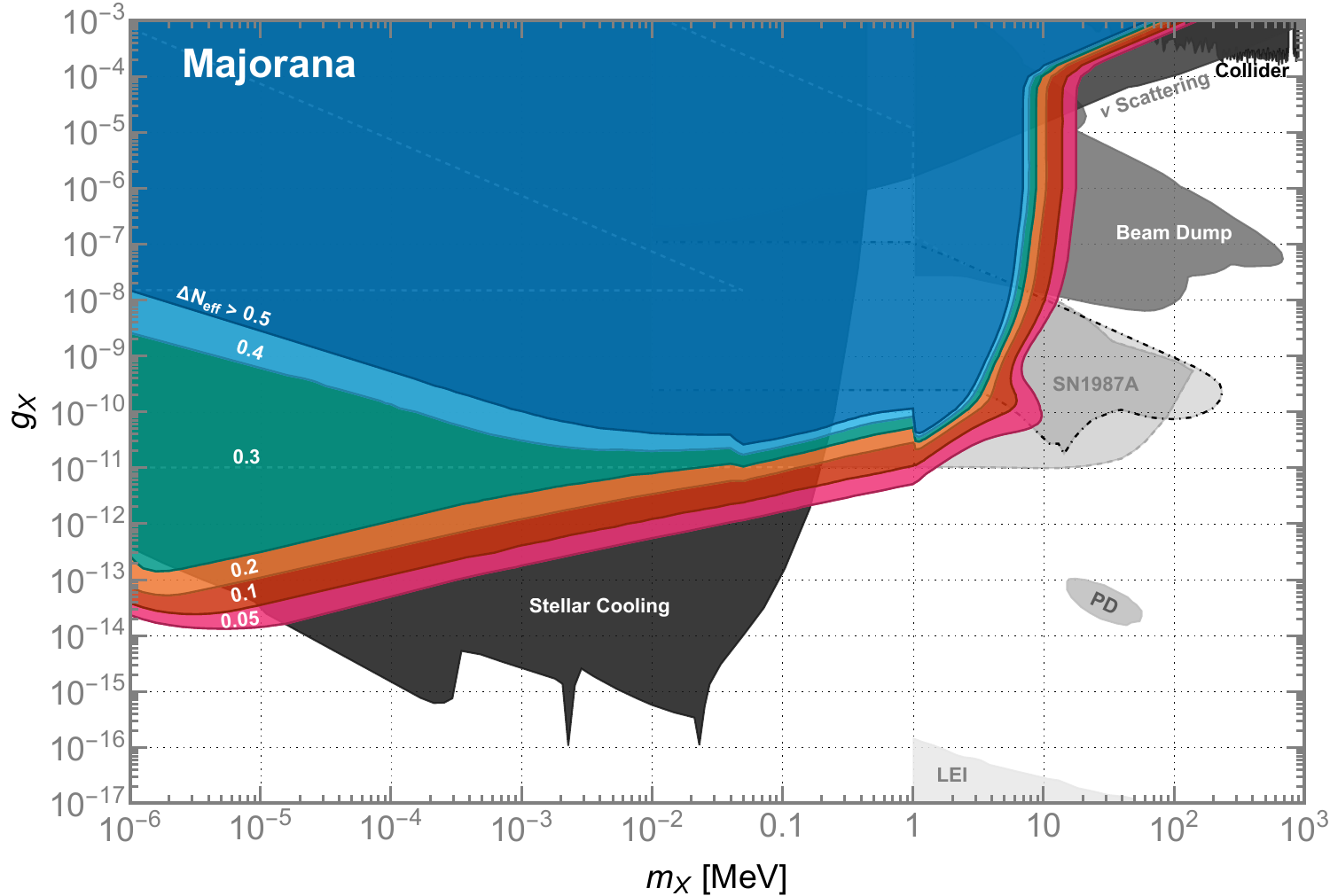}
\end{center}
\caption{Calculations of $\dneff$ for the ``Majorana'' neutrino case.
  The regions correspond to $\Delta N_{\rm eff} >$
  $0.05$ (pink),
  $0.1$ (red), 
  $0.2$ (orange),
  $0.3$ (green),
  $0.4$ (light blue),
  and $0.5$ (dark blue).
  The dashed gray and dot-dashed
  black lines show the constraints from SN1987A from \cite{Shin:2021bvz}
  and \cite{Croon:2020lrf} respectively. See section~\ref{otherCnstrntsSec}
  for a discussion of SN1987A and other background constraints.
  Current constraints from Planck data exclude $\dneff \gtrsim 0.3$-$0.4$
  (depending on the choice of dataset) \cite{Planck:2018vyg},
  while future CMB observatories, such as CMB-S4, are expected to reach
  $\dneff \sim 0.06$ \cite{Abazajian:2019eic}.
  See section~\ref{sec:constraintscurrentfuture}
  for further discussion of current and future constraints.
}
\label{dneffMFig}
\end{figure}
\begin{figure}[t]
\begin{center}
\includegraphics[width=1.0\linewidth]{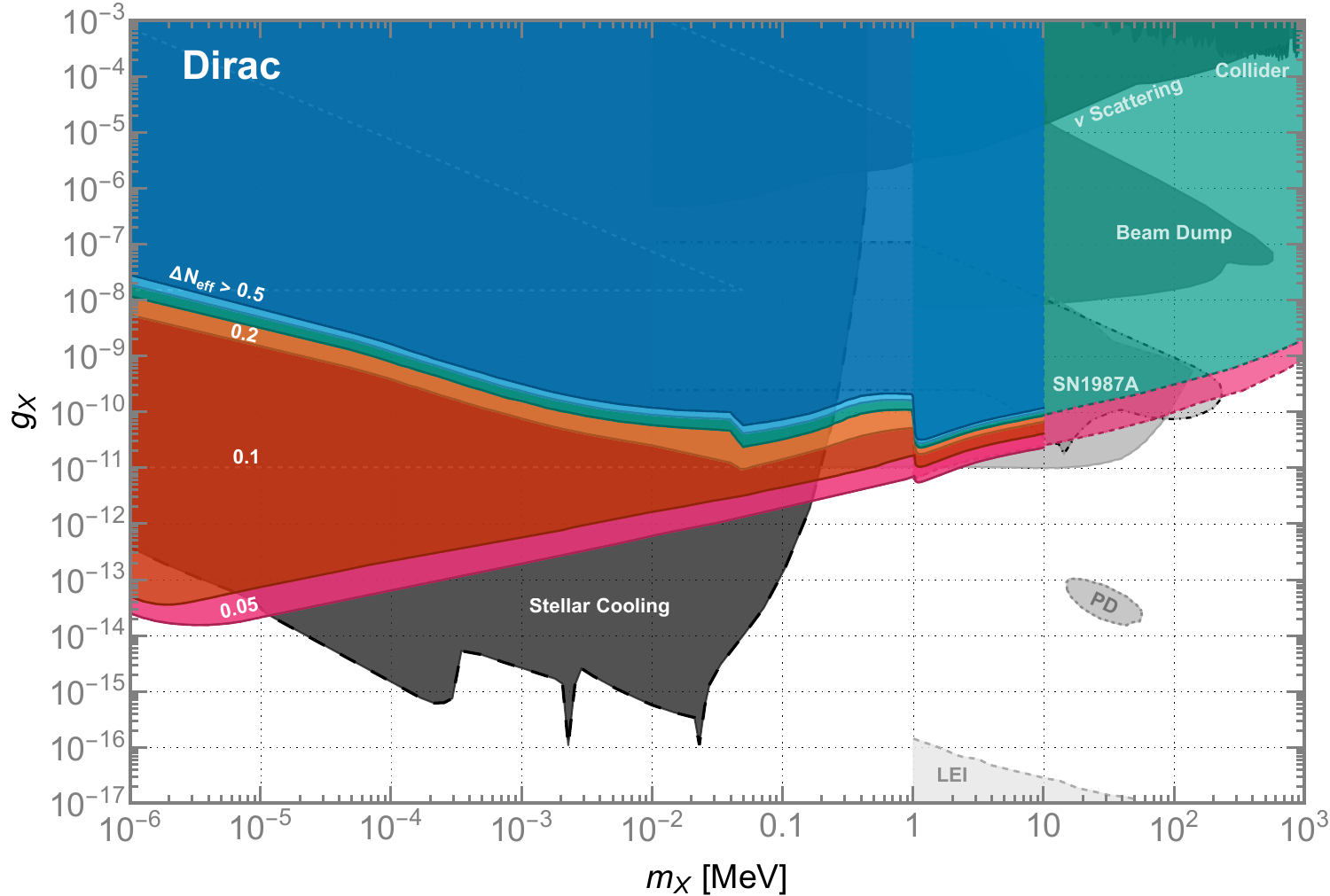}
\end{center}
\caption{Same as figure~\ref{dneffMFig}, but for the ``Dirac'' neutrino case.
  As before, the regions correspond to $\Delta N_{\rm eff} >$
  $0.05$ (pink),
  $0.1$ (red), 
  $0.2$ (orange),
  $0.3$ (green),
  $0.4$ (light blue),
  and $0.5$ (dark blue).
  The dashed pink and green contours correspond to $\dneff= 0.06 \,\{0.3\}$
  in \cite{Adshead:2022ovo} respectively. The dashed gray and dot-dashed 
  black lines show the constraints from SN1987A from \cite{Shin:2021bvz}
  and \cite{Croon:2020lrf} respectively. See section~\ref{otherCnstrntsSec}
  for a discussion of SN1987A and other background constraints.
}
\label{dneffDFig}
\end{figure}
We now show the full results of our evaluation of $\neff$ in the  
$(m_X, g_X)$ plane in figure~\ref{dneffMFig}
(Majorana neutrino case) and figure~\ref{dneffDFig}
(Dirac neutrino case).

Let's first discuss the Majorana neutrino case, figure~\ref{dneffMFig}.
This is a contour plot in $\dneff$, where colored regions provide
our result for $\dneff$ as shown in the plot.  In each colored
region, $\dneff$ is greater than the value as shown,
and less than its nearest neighboring region.  For the
dark blue region, $\dneff > 0.5$, and we did not 
delineate any larger values, for reasons that we will
discuss shortly.

There are three regimes of $m_X$ that reveal the qualitatively
different processes that are occurring to determine the contribution
to $\dneff$.  In the low mass region, $m_X < 2 m_e$, 
with $\dneff \sim 0.05$ to $0.2$, the process
$X \leftrightarrow \nu_L\bar{\nu}_L$ is approaching thermal
equilibrium.
The shape of the contour in $(m_X, g_X)$ space,
$g_X \sim \sqrt{(24\pi m_X)/M_{\text{Pl}}}$, 
can be obtained from $\Gamma_{X \lra \nunuL}/H \sim 1$
and $T = m_X$.  The precise value of
$\neff$ requires the numerical modeling of the
Boltzmann evolution.  (However, as we will see in
section~\ref{sec:semianalytic},
we can also estimate the value of $\dneff$ using
semi-analytic arguments.)
We see that in this region, the bounds on $g_X$
strengthen with \emph{decreasing} $m_X$.
Holding $m_X$ fixed, as $g_X$ is further increased,
$X$ and left-handed neutrinos reach thermal equilibrium,
saturating the contribution to $\dneff \sim 0.3$.
Once the process $X \leftrightarrow \nu_L\bar{\nu}_L$ is
in thermal equilibrium, there is no further increase in $\dneff$
for larger values of $g_X$, as evident by the green plateau region
in figure~\ref{dneffMFig}. 
The $\dneff = 0.4$,$0.5$ contours, that have a different shape,
arise because there are
additional scattering processes that are contributing to $\dneff$:
the $2 \leftrightarrow 2$ processes $\ee \ra X\gamma$ and
$e \gamma \ra eX$.  These processes become efficient when $n_e \langle \sigma v \rangle _{2 \leftrightarrow 2}/ H\sim 1$ at $T\sim m_e$; this corresponds to couplings of order $g_X \gtrsim 10^{-10}$ if we naively treat the rates in vacuum. However, unlike $X \leftrightarrow \nu\bar{\nu}$,
these processes have important finite temperature
corrections\footnote{Had finite temperature effects been neglected,
  the vacuum $\ee \ra X\gamma$ and $e \gamma \ra eX$ rates are $m_X$
  independent (for $m_X\ll m_e$), and so the $0.4$,$0.5$ contours would
  have become horizontal.}
that require the use of the in-medium coupling $g_{X,m}$
that is suppressed in the small $m_X$ limit as explained in
appendix~\ref{app:finiteTemperatureEffects}.
The small kink in these contours near $m_X \sim 0.05$~MeV is the result
of a linear interpolation as discussed in section \ref{eXSec}.

In the intermediate mass region,
$1 \; {\rm MeV} \; \lesssim m_X \lesssim 20 \; {\rm MeV}$,
the contours in $\dneff$
rise rapidly in $g_X$ as $m_X$ increases.
The major contributor to the rapid rise of $\dneff$ contours
in $g_X$ as $m_X$ increases is that as $X$ becomes more massive,
it becomes nonrelativistic earlier in the universe, and most of its
decays back into SM matter occurs before neutrino decoupling.
The SM processes rapidly equilibrate the relativistic
species, erasing the effect of $X$ production and decay.  
There is detailed structure, such as the shape of the contours
near $\dneff \sim 0.05$ and masses $m_X \sim 10$~MeV, 
that is a more complicated interplay between $X$,
electrons, and neutrinos.  

Finally, in high mass region,
$m_X > 20$~MeV, the Boltzmann evolution is calculated only
involving photons, electrons, and neutrinos, with
virtual $X$-exchange in the $2 \lra 2$ processes 
$e^{+}e^{-}\lra\bar{\nu}_\alpha\nu_\alpha$,
$e^{\pm}\nu_\alpha \lra e^{\pm}\nu_\alpha$, and
$e^{\pm}\bar{\nu}_\alpha \lra e^{\pm}\bar{\nu}_\alpha$.
The $\dneff$ contours in this region have a shape that
can be determined by requiring $X$-exchange is
no larger than $W$, $Z$ exchange.  This occurs
when $g_X \sim m_X \sqrt{G_F}$, providing an excellent
characterization of the shape.
The physics of this process is that if $X$ exchange
exceeds the weak interaction rate, this delays
neutrino decoupling in the evolution of the universe,
which has the effect of increasing $\dneff$.
Our contour shape agrees with that shown
in \cite{Knapen:2017xzo}. 

Now, let's discuss the contours in the Dirac neutrino
case, figure~\ref{dneffDFig}, comparing and contrasting with
the Majorana neutrino case.
In the low mass region, $m_X < 2 m_e$, 
the shape of the $\dneff$ contours is the same as the
Majorana case, with the only difference being that
the contour values of $\dneff$ are somewhat lower.
As we discussed in section~\ref{sec:evolutionboltzmann}, what is happening is
that the the energy density of primordial left-handed neutrinos
is now distributed among $X$, $\nuL$, and $\nuR$.
Hence, the maximum energy density that $X$ can achieve is smaller
than in the Majorana case, and so the entropy dump of $X$ into
neutrinos is smaller.  Later in section~\ref{sec:semianalytic}
we will verify this with a semi-analytic analysis. To be explicit, the  $\dneff = 0.05$ contour corresponds to the $X\lra \nunuL$ and $X\lra \nunuR$ processes approaching equilibrium. Thermal equilibrium of these processes occurs around $\dneff  \sim 0.1$ at which point the contribution to $\dneff$ from left and right handed neutrinos is saturated. We also see that at larger couplings the constraint from
$\ee \ra X\gamma$ and $e \gamma \ra eX$ processes remains,
but now extends to the $0.2$-$0.5$ contours analogous to
the $0.4$,$0.5$ contours described in the Majorana case.

At intermediate masses, $m_X \gtrsim 1$~MeV, as $X$ is populated,
then becomes nonrelativistic, it decays into both left-handed \emph{and}
right-handed neutrinos.  Hence, it contributes significantly to
$\Delta N_{\rm eff}$ because $X$ is providing a mechanism to siphon
off energy density from left-handed neutrinos into a ``new'' species,
right-handed neutrinos, that are completely decoupled from the SM\@.
This is unlike the Majorana case, where again for
$m_X \gtrsim ~{\rm few}$~MeV, $X$ decays back entirely into SM states
that equilibrate with the SM through electromagnetic or weak interactions.
This effect of populating right-handed neutrinos implies the
contours in $\dneff$ are at much smaller values of $g_X$
when $m_X \gg 1$~MeV\@.  We have calculated the result to
$m_X = 10$~MeV, and then we show the contours obtained by
\cite{Adshead:2022ovo} for larger masses.  In particular,
the dashed pink and green contours, that correspond to
$\dneff= 0.06 \,\{0.3\}$, match well onto our calculations
for $m_X < 10$~MeV\@.

There is one last issue to discuss.  At the very smallest masses
that we consider, the lifetime of $X$ becomes comparable to
the time of recombination
\begin{eqnarray}
  \tau = \frac{c}{\Gamma}
  &\simeq& 330 \; {\rm kyr} \times
           \left( \frac{1 \; {\rm eV}}{m_X} \right)
           \left( \frac{4 \times 10^{-14}}{g_X} \right)^2 \, ,
\label{eq:Xlifetime}
\end{eqnarray}
for the Majorana neutrino case (for the Dirac case
the lifetime is $1/2$ of this). 
This means not all of the $X$ bosons that freeze-in have
decayed back to relativistic neutrino degrees of freedom.
We have taken this effect into account by evaluating
$\neff$ after all of the $X$ bosons have decayed or
at the temperature of recombination,
$T \simeq 0.3$~eV, whichever occurs first.  This can be seen in both
Figs.~\ref{dneffMFig},\ref{dneffDFig}, where the contours
of $\dneff$ change slope from positive to negative as
$m_X$ drops below $\sim {\rm few} \; {\rm eV}$,
leading to slightly weaker bounds on $g_X$
for the smallest masses of $X$ that we consider.
A very similar effect was also found in \cite{Sandner:2023ptm}
for the $U(1)_{\mu - \tau}$ model.

\subsection{Other constraints}
\label{otherCnstrntsSec}

In Figs.~\ref{dneffMFig},\ref{dneffDFig}, we have also shown several
other constraints on the mass and coupling of a $U(1)_{B-L}$
gauge boson.  An overview of previously obtained constraints
can be found from \cite{Harnik:2012ni, Heeck:2014zfa, Bauer:2018onh, Ilten:2018crw, Coffey:2020oir}.

Supernova 1987A (SN1987A) provides well-known constraints on
the emission of light mediators \cite{Raffelt:1987yt}.
The presence of a light vector mediator provides an additional cooling
mechanism in the proto-neutron star that competes with electroweak
interactions and alters its neutrino luminosity.
Constraints on a variety of gauged $U(1)$ bosons have been derived
in \cite{Chang:2016ntp, Hardy:2016kme, Croon:2020lrf, Shin:2021bvz}.
The constraints for $U(1)_{B-L}$
in \cite{Croon:2020lrf} and \cite{Shin:2021bvz} are
derived using slightly different methods and arrive at slightly
different constraints.  For completeness, we show both of these
in the figures, and we will return to this issue in future work. 

Similarly, constraints can be set if the energy loss rate due to
light gauge boson emission from the sun, red giants (RG), and
horizontal branch (HB) stars exceeds a maximally allowed limit
\cite{An:2014twa, Hardy:2016kme, Li:2023vpv}. Note both constraints
from stellar cooling and SN1987A are derived explicitly for the case
where right-handed neutrinos are heavy and do no contribute to the
dynamics (``Majorana case''). To the best of our knowledge,
a similar derivation has not been done in the literature for the
Dirac case, though we do not suspect major differences in the
results.  So, we have included these constraints in figure~\ref{dneffDFig},
though denoted as dashed to remind the reader of this subtlety.

We also show the constraints from photodissociation (PD)
of light elements, and separately, late energy injection (LEI) at
the CMB era, that were obtained for $U(1)_{B-L}$ in \cite{Coffey:2020oir}. 
These constraints are at much smaller coupling,
where $X$ was very far from reaching thermal equilibrium,
and for masses $m_X > 2 m_e$ so that the decay
$X \ra e^+e^-$ is kinematically available.  
In Figs.~\ref{dneffMFig},\ref{dneffDFig} we show only the upper part
of the LEI contour; we refer readers to \cite{Coffey:2020oir}
for the full region that extends to couplings values $g_X < 10^{-17}$. 

Neutrino experiments that probe electron-neutrino or
coherent elastic neutrino-nucleus scattering (CE$\nu$NS) are sensitive
to the contribution of $X$ to their amplitudes
\cite{TEXONO:2009knm, Bilmis:2015lja, Borexino:2017rsf, Lindner:2018kjo, COHERENT:2020iec, COHERENT:2021xmm, A:2022acy, Coloma:2022umy}. 
Experiments searching decays into invisible final states can also be used
to set constraints with the assumption that invisible states can be
treated as decays to neutrino final states
\cite{Fox:2011fx, BaBar:2017zmc, NA64:2017vtt, NA62:2019meo, Belle-II:2019qfb}.
In beam dump experiments \cite{Bernardi:1985ny, CHARM:1985anb, Konaka:1986cb, Riordan:1987aw, Bjorken:1988as, Bross:1989mp, Davier:1989wz, NOMAD:2001eyx, Blumlein:2011mv, Gninenko:2011uv, Gninenko:2012eq},
$X$ is produced via the bremsstrahlung process where a beam of
electron/protons is dumped on some target material; $X$ then travels
through shielding material before decaying into SM particles that
can be observed in the detector. Constraints are derived by comparing
the number of expected and observed events in the detector.
Finally, colliders set constraints on promptly decaying
$X$ \cite{BaBar:2014zli, LHCb:2017trq, Ilten:2018crw}.
These beam dump and collider constraints apply to both the
Majorana and Dirac case.
However in the Dirac case, collider and beam dump and neutrino scattering
constraints have been recast to account for the additional branching fraction
of $X\rightarrow \nunuR$ \cite{Heeck:2014zfa}.


\subsection{Constraints from existing CMB observations
  and prospects for future observatories}
\label{sec:constraintscurrentfuture}

We have presented our results in Figs.~\ref{dneffMFig},\ref{dneffDFig}
with our calculated contours of $\dneff$ as would be measured by the CMB
within the $(m_X, g_X)$ model parameter space.  Determining the current
and future constraints on $\dneff$ requires significant care with regard
to two critical issues.
First, the \emph{era} at which new physics modifies $\neff$,
and what other cosmological parameters are also modified,
particularly between BBN and CMB\@.  Second, the \emph{datasets}
used to constrain the cosmological parameters, and the potential
correlations among observables.  In this section, our focus is on
the CMB determination of $\neff$.  

Nevertheless, we can consider two proxies for BBN constraints
on new physics that are potentially different from $\neff$.
One is the helium mass fraction $Y_p$, that we consider in
detail in section~\ref{bbnSec}.  A second quantity is
the baryon-to-photon ratio, $\eta$.  
In general, new physics can contribute to a shifts in
$\eta^{CMB}$ relative to $\eta^{BBN}$
when, for example, there is a period
of late energy injection into photons well after
the temperature of helium synthesis, $T \sim 0.07$~MeV
\cite{Berlin:2019pbq,Yeh:2022heq}.
In the SM, there is negligible energy injection into
photons at this temperature because $\ee$ annihilation is
virtually complete, and so $\eta^{CMB} \simeq \eta^{BBN}$.
Considering $U(1)_{B-L}$, we also do not have late energy
injection directly into photons.  We do have energy injection
into $\ee$ from $X \ra \ee$, but since $\ee$ annihilation
did not affect $\eta$ in the SM, we do not expect these
contributions from $X \ra \ee$ to shift $\eta$ from BBN
to the CMB era either.

Focusing on $N_{\rm eff}^{\rm CMB}$, we can now turn to
the current constraints and future prospects for cosmological measurements
of $N_{\rm eff}^{\rm CMB}$.  The Planck collaboration (PCP18)
have set the best bounds thus far, with various combinations of
datasets \cite{Planck:2018nkj, Planck:2018vyg}:
\begin{subequations}
\label{planckNeffEqns}
 \begin{align}
 \label{pcp1pEq}
\neff\cmb\vert^{\text{PCP18}}_{1\, \text{pe}}&=2.99\,^{+0.34}_{-0.33} \;&&( 95\%,\, \text{TT, TE, EE+lowE+ lensing+BAO})\\
 \label{pcp2pEq}
\neff\cmb\vert^{\text{PCP18}}_{+Y_p}&=2.99\,^{+0.43}_{-0.40} \;&&( 95\%,\, \text{TT, TE, EE+lowE+ lensing+BAO+Aver \cite{Aver:2015iza}})\\
\neff\cmb\vert^{\text{PCP18}}_{+H_0}&=3.27\,^{+0.30}_{-0.30} \;&&( 95\%,\, \text{TT, TE, EE+lowE+ lensing+BAO+R18 \cite{Riess:2018uxu}}).
 \label{pcp3pEq}
 \end{align}
\end{subequations}
The first value, $\neff\cmb\vert^{\text{PCP18}}_{1\, \text{pe}}$, is the on 1-parameter extension to the base-$\Lambda \text{CDM}$ model six parameter fit. However, the primordial helium mass fraction $Y_p$ and $\neff$ are partially degenerate as they both affect the CMB damping tail (i.e. \cite{Hou:2011ec}). Allowing $Y_p$ to vary with $\neff$ gives a value of $\neff\cmb\vert^{\text{PCP18}}_{+Y_p}$ and $Y_p^{\text{BBN}}\vert^{\text{PCP18}}_{+Y_p}=0.2437\,\pm 0.0080$ at $95\%$ C.L.. The combined fit eases the constraint on both $\neff$ and $Y_p$ which has a PDG recommended value of $Y_p\vert^{\text{PDG}}=0.245 \,\pm 0.006$ at $95\%$ C.L.~\cite{Workman:2022ynf} based on the analyses in \cite{Aver:2020fon,Valerdi:2019beb,Fernandez:2019hds,Kurichin:2021ppm,Hsyu:2020uqb}. Finally,  the $\neff\cmb\vert^{\text{PCP18}}_{+H_0}$ partially addresses the $\sim 3 \sigma$ tension in the Hubble constant, $H_0$, between CMB and local observation determinations \cite{Riess:2018uxu,Riess:2016jrr,Riess:2018byc} by including the analysis in R18 \cite{Riess:2018uxu}. This fit could potentially increase the tension with weak galaxy lensing and (possibly) cluster count data since it requires an increase in the $\sigma_8$ and a decrease in $\Omega_m$ values.

After the PCP18 data release and analysis, Yeh2022 \cite{Yeh:2022heq} presents independent BBN and CMB limits on $\neff$ and $\eta$ using likelihood analyses and updated evaluations for nuclear rates. The BBN likelihood functions are obtained by varying nuclear reaction rates within their uncertainties via a Monte Carlo, while the CMB likelihoods are derived from PCP18 and marginalized over $Y_p$. With various combinations of datasets,
\begin{subequations}
\label{Yeh2022NeffEqns}
 \begin{align}
 \label{firstYehEqn}
 &\neff\bbn\vert^{\text{Yeh2022}}_{\text{BBN only}}&&=2.889\,\pm\, 0.229\;&&( 68.3\%, Y_p + \text{D})\\
  \label{secondYehEqn}
&\neff\cmb\vert^{\text{Yeh2022}}_{\text{CMB only}}&&=2.800\,\pm\, 0.294\;&&( 68.3\%, \text{TT, TE, EE+lowE+ lensing} )\\
&\neff\vert^{\text{Yeh2022}}_{\text{BBN+CMB}}&&=2.898\,\pm\, 0.141\;&&( 68.3\%,\text{TT, TE, EE+lowE+ lensing}+ Y_p + \text{D}).
\label{thirdYehEqn}
\end{align}
\end{subequations}
The first value, $\neff\bbn\vert^{\text{Yeh2022}}_{\text{BBN only}}$, shows limits on $\neff$ at BBN where the BBN likelihood is convolved with the observation determination of $Y_p$ and deuterium abundance D \cite{Pettini:2012ph,Cooke:2013cba,Riemer-Sorensen:2014aoa,Cooke:2016rky,Balashev:2015hoe,Riemer-Sorensen:2017pey,Zavarygin:2018ara,Cooke:2017cwo} likelihoods. The second value, $\neff\cmb\vert^{\text{Yeh2022}}_{\text{CMB only}}$, differs from equation~(\ref{pcp1pEq}) as the Yeh2022 analysis does not assume any relation between the $^4\text{He}$ abundance and the baryon density.
The third value is the limit set by combining BBN and the CMB assuming no new physics after nucleosynthesis, however this value is not applicable here since $X$ interactions can affect $\dneff$ between BBN and CMB\@.

Comparing these observations to our results,
Figs.~\ref{dneffMFig},\ref{dneffDFig}, 
a conservative analysis suggests $\dneff^{\rm CMB} \lesssim 0.3$-$0.4$
to 95\% C.L., and hence the regions labeled dark and light blue
are ruled out, with the green region strongly disfavored.
The 2018 Planck analysis that incorporated the local Hubble measurement
into the fit yielded $\neff^{\rm CMB} \simeq 3.27$, Eq.~(\ref{pcp3pEq}),
which would favor the orange region, however this is in tension with
more recent analyses of the CMB data.
If the ultimate resolution of the Hubble tension involves a slightly
larger value of 
equation~(\ref{pcp3pEq}), then the orange region is favored.  
Future observatories, including
SPT-3G \cite{SPT-3G:2014dbx}, 
CORE \cite{CORE:2016npo}, 
Simons Observatory \cite{SimonsObservatory:2018koc}, 
PICO \cite{NASAPICO:2019thw},
CMB-S4 \cite{Abazajian:2019eic},
CMB-HD \cite{Sehgal:2019ewc},
are anticipated to reach much smaller values, 
for example CMB-S4 is anticipated to ultimately obtain $\dneff \simeq 0.06$
at 95\% C.L. \cite{Abazajian:2019eic}.
Here we see these future experiments are capable of probing
a significantly larger fraction of the $(m_X, g_X)$ parameter space
in both the Majorana and Dirac neutrino cases.  
The future CMB observations will be the most sensitive to
the presence of a gauged $U(1)_{B-L}$ boson 
in the regions $1 \; {\rm eV} \lesssim m_X \lesssim 20 \; {\rm eV}$,
$0.2 \; {\rm MeV} \lesssim m_X \lesssim 20 \; {\rm MeV}$,
and in the Dirac case, $m_X \gtrsim 200$~MeV \cite{Adshead:2022ovo}.


\section[Semi-analytic estimates of delta \texorpdfstring{$\dneff$}{Neff}]{Semi-analytic estimates of $\dneff$}
\label{sec:semianalytic}

There is an interesting way to cross-check our results in the
regime $m_X \ll m_e$, where the only interactions of $X$ are
with neutrinos.  This semi-analytic approach involves two steps.
First, we calculate the energy densities and number densities of $X$,
assuming thermalization occurs at a temperature $T > m_X$.
For the purposes of this estimate, we assume that thermalization occurs
when $X$ is relativistic and can be treated as massless.
The key nontrivial part of this first step is that we are solving
for not only the energy densities but also the chemical potentials
as $X$ is frozen-in to equilibrium with neutrinos.  Next, we use
entropy conservation to determine the net effect of $X$ decaying
back into neutrinos.  Again, this step is nontrivial because we also
must take into account the nonzero chemical potentials of $X$
and the various neutrino species that (were) in equilibrium.

Given these assumptions, we are able to semi-analytically calculate
the contribution to $\Delta N_{\rm eff}$. 
We now present the calculation in two regimes:  
first, the Majorana case, to get a clear sense of the method
and its result.  Next, the Dirac case, however we will generalize
this to include an arbitrary number of additional species of
(right-handed) fermions that carry lepton number.  This case will
allow us to provide semi-analytic estimates of $\Delta N_{\rm eff}$
in a wider range of models that may be relevant for dark matter
and cosmology.

\subsection{Majorana case}
\label{sec:majoranacasesemianalytic}
  
An initial density of left-handed neutrinos with temperature $T_\nuL$
is converted to a density of left-handed neutrinos and $X$,
\begin{subequations}
\label{thermEntropy}
\begin{eqnarray} 
\label{denEqualityEntropyM}
  \rho_\nuL(T_\nuL,0) &=& \rho_\nuL(T_{\text{eq}},\mu_{\text{eq}})
                          + \rho_X(T_{\text{eq}},2\mu_{\text{eq}}) \\ 
  n_\nuL(T_\nuL,0) &=& n_\nuL(T_{\text{eq}},\mu_{\text{eq}})
                       + 2n_X(T_{\text{eq}},2\mu_{\text{eq}}) ,
\end{eqnarray}
\end{subequations}
with $T_{\text{eq}},\mu_{\text{eq}}$ thermal equilibrium temperature and chemical potential. In the second equation, the factor of $2$ arises because one $X$ is produced at the expense of two neutrinos in $\nunuiL \rightarrow X$. These conditions give,
\begin{equation}
\label{eqTempM}
T_{\text{eq}} = 1.208 \,T_\nuL \; , 
\quad\quad
\mu_{\text{eq}} = -1.166 \,T_\nuL  
\end{equation}
which gives a maximum density ratio $\varrho_X\equiv\rho_X/(\rho_\nuL+\rho_X)  = 0.1642$. Had we neglected chemical potential, equation~(\ref{denEqualityEntropyM}) becomes a degree of freedom counting argument which yields $\varrho_X=3/(6\times \frac{7}{8}+3)=\frac{4}{11} = 0.3636$ significantly overestimating the maximum $X$ abundance. Instead, the chemical potential appears as $\varrho_X= 3\LiF(x)/[6\LiF(-x)+3\LiF(x)]$ with $x\equiv e^{\mu/T}$ and $\LiF$ is the $4^{\text{th}}$ order polylogarithm.

After thermal equilibrium is reached, we use entropy conservation to write,
\begin{subequations}
\label{entropyAfterDecayM}
\begin{eqnarray} 
  \left[ s_\nuL(T_{\text{eq}},\mu_{\text{eq}})
  + s_X(T_{\text{eq}},2\mu_{\text{eq}})\right] a'^3
  &=& s_\nuL(T_\nu,\mu_\nu)a^3 \\
  \left[n_\nuL(T_{\text{eq}},\mu_{\text{eq}})
  + 2n_X(T_{\text{eq}},2\mu_{\text{eq}}) \right] a'^3
  &=& n_\nuL(T_\nu,\mu_\nu)a^3.
\end{eqnarray}
\end{subequations}
The left-hand side starts at scale factor $a'$ with equilibrium abundance of all relevant particles, while the right-hand side ends at a scale factor $a$ with $T_\nu \ll m_X$, where the entire distribution of $X$ bosons has decayed into neutrinos. We can relate the scale factors through $T_\gamma' a'=T_\gamma a$, since the photon plasma is decoupled from neutrinos. Solving equation~(\ref{entropyAfterDecayM}) gives,
 \begin{equation}
\label{TgammaTnuAfterDecayEntropyM}
T_\gamma/T_\nu = 1.278 \, ,
\quad\quad
T_\nu/\mu_\nu = - 3.486
\end{equation}
which can be used to calculate
 \begin{equation}
\label{entropyNeffM}
\dneff=\neff-\neff^{\text{SM}}=0.25 \, .
\end{equation}
As we discussed in section~\ref{cnstrntSec}, the diagonal contours of
$\dneff$
in the light regime $m_X \ll 2 m_e$ correspond to when the
$1 \lra 2$ process $X \lra \nu\bar{\nu}$ approximately reaches
thermal equilibrium.  The highest contour shown is $0.3$,
that is quite close to our semi-analytic estimate in equation~(\ref{entropyNeffM}).
Of course the semi-analytic method required certain
approximations to be taken.
Namely, in solving equation~(\ref{entropyAfterDecayM}) we used the SM value
of $T_\gamma/T_\nu$ in equation~(\ref{NeffinSM}) to relate the scale factors
$T_\gamma' a' = T_\gamma a$; and using the numerical value obtaining
by solving the Boltzmann equations
in equation~(\ref{sevenboltzmannGeneral}) reproduces the correct value
of $\dneff$.

\subsection{Dirac and generalized cases}

We now redo the analysis of section~\ref{sec:majoranacasesemianalytic}
by adding into the thermal bath an arbitrary number $N_R$
of (Weyl) fermion species.\footnote{Here $N_R$ refers to
  the number of $L = \pm 1$ particles, and so not necessarily
  are they are strictly ``right-handed'' with $L=1$. 
  The Dirac case has $N_R = 3$ and $L = 1$ for all species.
  The sign of the lepton number of any of these species
  does not enter our discussion below.}
As a slight abuse of notation,
we will refer to these fermions as ``right-handed neutrinos'' below,
but it should be understood that these fermion fields
have $L = \pm 1$ and may or may not
pair up with the left-handed neutrinos to gain Dirac masses.
All of these right-handed neutrinos will be considered
to be either (i) massless, or (ii) having a mass
$m_{\nu_R} \lesssim m_X/2$.  
The Dirac case follows as a special case of this generalization
where $N_R = 3$, $m_{\nu_R} = 0$, and two (or three) of these
fermion species pair up with left-handed neutrinos.

Following Eqs.~(\ref{thermEntropy}), an initial density of
left-handed neutrinos with temperature $T_\nuL$
is converted to a energy (number) density of left- and right-handed neutrinos
[denote collectively by $\rho_\nu$ ($n_\nu$)]\footnote{Note that both quantities $\rho_\nu$ and $n_\nu$ include a factor of $g_\star = 6$ corresponding to three fermionic Weyl degrees of freedom in their definition.}
and $X$,
\begin{subequations}
\label{thermEntropygeneralized}
\begin{eqnarray}
\label{denEqualityEntropygeneralized}
  \rho_\nuL(T_\nuL,0) &=& \left( 1 + \frac{N_R}{3} \right)
                          \rho_\nu(T_{\text{eq}},\mu_{\text{eq}})
      + \rho_X(T_{\text{eq}},2\mu_{\text{eq}}) \\
  n_\nuL(T_\nuL,0) &=& \left( 1 + \frac{N_R}{3} \right)
                       n_\nu(T_{\text{eq}},\mu_{\text{eq}})
      + 2 n_X(T_{\text{eq}},2\mu_{\text{eq}})
\end{eqnarray}
\end{subequations}
with $T_{\text{eq}},\mu_{\text{eq}}$ thermal equilibrium temperature
and chemical potential.

After thermal equilibrium is reached, we can again utilize entropy
conservation to write,
\begin{subequations}
\label{entropyAfterDecaygeneralized}
\begin{eqnarray}
  \left[
  \left( 1 + \frac{N_R}{3} \right) s_\nu(T_{\text{eq}},\mu_{\text{eq}})
  + s_X(T_{\text{eq}},2\mu_{\text{eq}})
  \right] a'^3
  &=& \left[ \left( 1 + \frac{N_R}{3} \right) s_\nu(T_\nu,\mu_\nu) \right]
      a^3 \\
  \left[
  \left( 1 + \frac{N_R}{3} \right) n_\nu(T_{\text{eq}},\mu_{\text{eq}})
  + 2 n_X(T_{\text{eq}},2\mu_{\text{eq}})
  \right] a'^3
  &=& \left[ \left( 1 + \frac{N_R}{3} \right) n_\nu(T_\nu,\mu_\nu)
      \right] a^3 \, .
\end{eqnarray}
\end{subequations}
Again, the left-hand side starts at scale factor $a'$ with
equilibrium abundance of all relevant particles and the right-hand side
ends at scale factor $a$ with $T_\nu \ll m_X$ where all
$X$ distribution has decayed into neutrinos.

\subsubsection{Dirac case}

In the Dirac case ($N_R = 3$; $m_{\nu_R} = 0$),
Eqs.~(\ref{thermEntropygeneralized}) give
\begin{equation}
\label{eqTempDirac}
T_{\text{eq}} = 1.1088 \,T_\nuL \, , 
\quad\quad  \mu_{\text{eq}} = -1.3485 \,T_\nuL
\quad\quad \mbox{(Dirac case)}
\end{equation}
which gives a maximum density ratio
$\varrho_X\equiv\rho_X/(\rho_\nuL+\rho_X+\rho_\nuR) = 0.07049$.
(Again, had we neglected chemical potential,
equation~(\ref{denEqualityEntropygeneralized})
becomes 
$\varrho_X= 3/(12 \times \frac{7}{8} +3) = \frac{2}{9} = 0.2222$, 
overestimating the maximum $X$ abundance.)
Solving equation~(\ref{entropyAfterDecaygeneralized}) gives,
\begin{equation}
\label{TgammaTnuAfterDecayEntropyDirac}
T_\gamma/T_\nu = 1.322 \, ,
\quad\quad 
T_\nu/\mu_\nu = - 1.093
\quad\quad \mbox{(Dirac case)}
\end{equation}
which can be used to calculate
\begin{equation}
\label{entropyNeffDirac}
\dneff = \neff-\neff^{\text{SM}} = 0.09 \, .
\end{equation}
Like our discussion above in the Majorana case,
in figure~\ref{dneffDFig} the highest contour consistent with
approximate thermalization of $X \lra \nu\bar{\nu}$ is $0.1$,  
that agrees with the semi-analytic estimate
in equation~(\ref{entropyNeffDirac}).

\subsubsection{Large number of massless species}

Now let's utilize the generalized results in
Eqs.~(\ref{thermEntropygeneralized}) and (\ref{entropyAfterDecaygeneralized})
to apply to the case where there is a large number
of massless species $N_R \gg 1$ that $X$ can decay into.
Once $X$ and the $1 + N_R/3$ neutrinos have thermalized,
it is clear from Eqs.~(\ref{thermEntropygeneralized})
that the energy and number density of $X$ is suppressed
by the same factor, $1/(1 + N_R/3)$.  Most of the energy and number
density of left-handed neutrinos is transferred into
the larger number of right-handed neutrinos.
Then, following Eqs.~(\ref{entropyAfterDecaygeneralized}),
the effect of the small number of $X$ gauge bosons is negligible.
That is, in the limit $N_R \gg 1$, there is negligible entropy
dumping of $X$ into neutrinos, and so $\Delta N_{\rm eff} \ra 0$.
For smaller values of $N_R$, 
we find, for example
$\dneff = 0.12$ ($N_R = 2$), 
$\dneff = 0.07$ ($N_R = 4$) and
$\dneff = 0.03$ ($N_R = 8$).
Here we do need to emphasize that while the number of
relativistic degrees of freedom nearly matches the SM,
the \emph{flavor} of these degrees of freedom is overwhelmingly
in the form of the (sterile) neutral fermion species carrying 
lepton number.

\subsubsection{Massive species}

An alternative scenario occurs if the $N_R$ species
are massive, and thus are able to annihilate through $X$
back into (massless) left-handed neutrinos species.
Here what happens is that the large entropy of the
right-handed neutrinos siphons off virtually all of the
entropy in left-handed neutrinos.  In the limit $N_R \gg 1$,
with $m_{\nu_R} \lesssim m_X/2$, the overwhelming majority
of the original energy density of left-handed neutrinos end up
as nonrelativistic, massive (sterile) neutral fermion species.
This leaves a negligible amount of left-handed neutrinos,
and so in this limit, $\Delta N_{\rm eff} \ra -3$,
i.e., $N_{\rm eff} \ra 0$.  
As a few examples where all $N_R$ species are massive, we
find 
$\dneff = -0.64$ ($N_R = 1$), 
$\dneff = -1.15$ ($N_R = 2$) and
$\dneff = -1.48$ ($N_R = 3$).
Notice that $N_R \ge 1$ is already completely excluded by
the Planck constraints on the CMB, i.e., equation~(\ref{planckNeffEqns}).

\section[BBN \texorpdfstring{$Y_p$}{Yp}]{BBN $Y_p$}
\label{bbnSec}

\begin{figure}[t]
\includegraphics[width=1.0\textwidth]{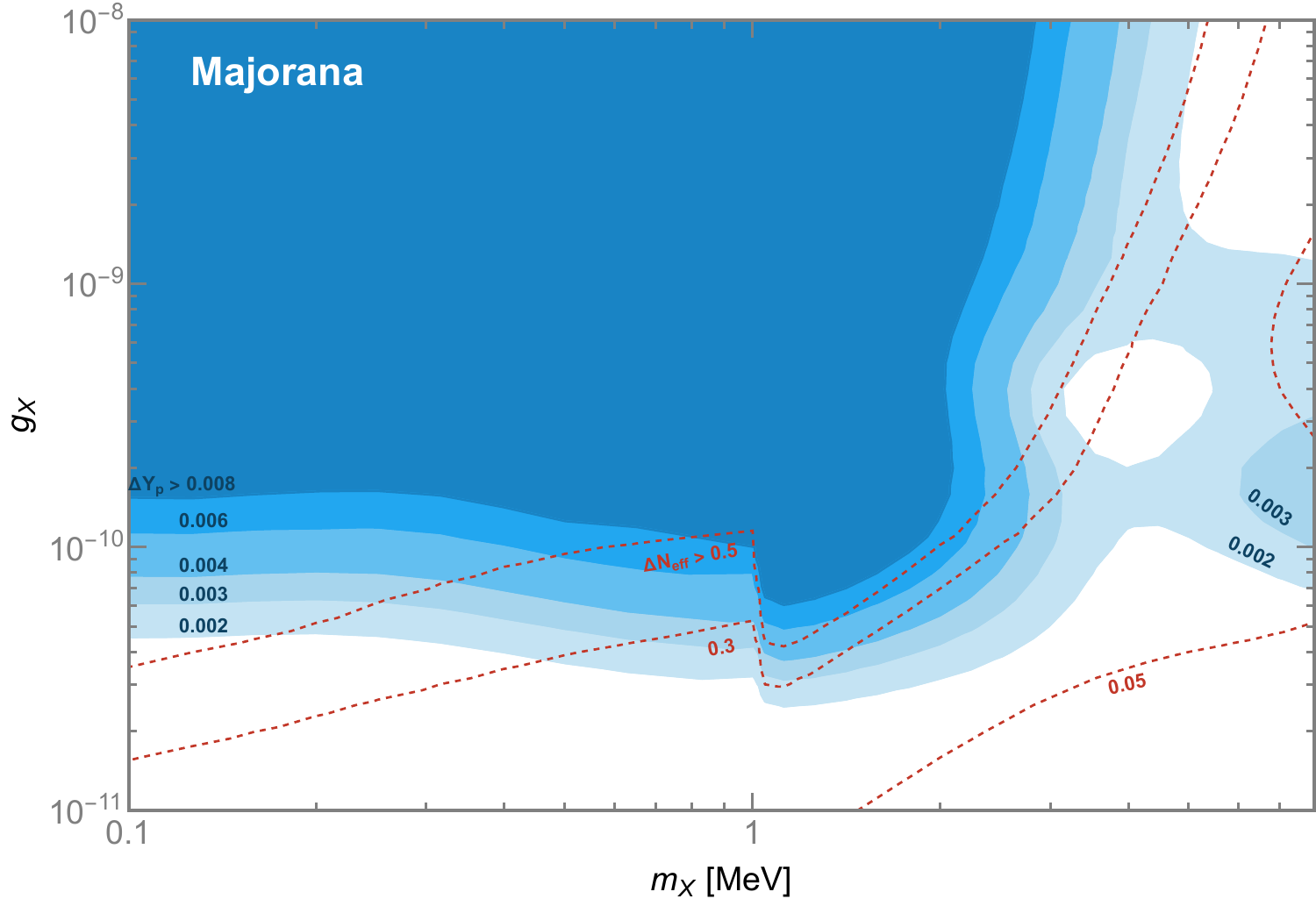}
\caption{$\Delta Y_p$ contours for ``Majorana'' 
  neutrino case.  The dashed red lines show the contours from our
  calculation of $\Delta N_{\rm eff}$ observed by the CMB,
  from figure~\ref{dneffMFig}.
  A conservative upper bound is $\Delta Y_p = 0.008$ at $95\%$ C.L. 
  \cite{Yeh:2022heq}, see section~\ref{bbnSec} 
  for further discussion.
}
\label{bbnCnstrntPltMFig}
\end{figure}
\begin{figure}[t]
\includegraphics[width=1.0\textwidth]{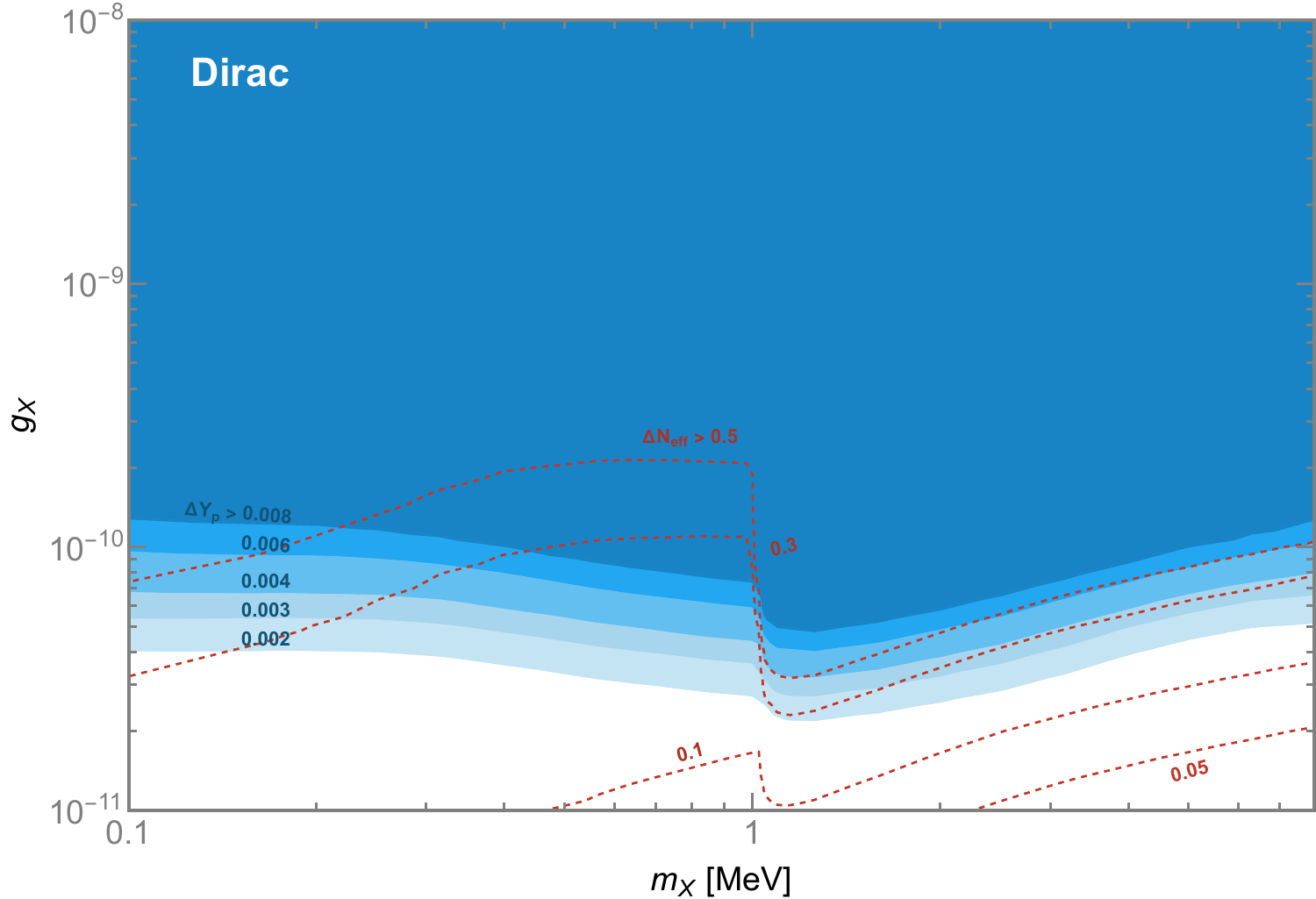}
\caption{Same as figure~\ref{bbnCnstrntPltMFig}, but for the ``Dirac'' 
  neutrino case.  The dashed red lines show the contours from our
  calculation of $\Delta N_{\rm eff}$ observed by the CMB,
  from figure~\ref{dneffDFig}.}
\label{bbnCnstrntPltDFig}
\end{figure}

The primordial abundances of nuclide species are critically sensitive
to the expansion history leading up to nucleosynthesis.
Precise determinations of these abundances requires solving Boltzmann
equations describing the density evolution of each nuclide species
$({}^{2}\text{H}, {}^{3}\text{He}, {}^{4}\text{He}...)$
(i.e. \cite{Iocco:2008va}). This can be done using publicly available
BBN code such as
\texttt{PArthENoPE} \cite{Consiglio:2017pot,Gariazzo:2021iiu}, \texttt{AlterBBN} \cite{Arbey:2011nf, Arbey:2018zfh},
\texttt{PRIMAT} \cite{Pitrou:2018cgg}, or \texttt{PRyMordial} \cite{Burns:2023sgx};
however, this is beyond the scope of this paper.

Instead, we follow an approximation employed in appendix A.4
of \cite{Escudero:2018mvt}, that involves modifying the neutron-to-proton
conversion rate to include a neutrino chemical potential following the
derivation in \cite{Dicus:1982bz,Dodelson:1992km}. The Boltzmann
equation describing the evolution of neutron fraction
$X_n=\frac{n_n}{n_n+n_p}$ is \cite{Sarkar:1995dd}:
\begin{subequations}
\label{npInteraction}
\begin{align} 
\frac{dX_n}{dt}&=\Gamma_{pn}(1-X_n) -\Gamma_{np} X_n,\\
\Gamma_{np}&=K \int_1^\infty d\epsilon \left[\frac{(\epsilon-q)^2(\epsilon^2-1)^{1/2}\epsilon}{(1+e^{-\epsilon z_\gamma})(1+e^{(\epsilon-q-\tilde{\mu}_\nu) z_\nu})}+ \frac{(\epsilon+q)^2(\epsilon^2-1)^{1/2}\epsilon}{(1+e^{\epsilon z_\gamma})(1+e^{-(\epsilon-q-\tilde{\mu}_\nu) z_\nu}}\right],\\
\Gamma_{pn}&=\Gamma_{np}(-q),
\end{align} 
\end{subequations}
where
$z_i=m_e/T_i$, $\tilde{\mu}_\nu=\mu_\nu/m_e$, $q=(m_n-m_p)/m_e$,
and $K\simeq (1.939 \tau_n)^{-1}$. We take $m_n-m_p=1.2933$~MeV and
neutron lifetime $\tau_n=878.4$~s \cite{Workman:2022ynf}.
Here $\Gamma_{np}$ is the rate for neutron to proton conversion 
through the weak interactions $n + \nu_e \lra p + e^-$,
$n + e^+ \lra p + \bar{\nu}_e$, and
$n \lra p + e^- + \bar{\nu}_e$. 
The evolution of $T_\gamma,\, T_\nu$ and $\mu_\nu$ are obtained separately,
by solving equation~(\ref{sevenboltzmannGeneral}) as described in
section~\ref{boltzSec}. To estimate the helium abundance $Y_p$,
we take the limit where all remaining neutrons form helium around
the temperature where photons no longer dissociate deuterium
$T_D=0.073$~MeV; in other words, 
\begin{equation}
\label{YpEqn}
Y_p\simeq 2X_n\vert_{T=T_D} \, .
\end{equation}

In the SM, applying this procedure yields
$Y_p\vert^{\text{equation } \ref{YpEqn}}_{\text{SM}}=0.248$
in agreement with $Y_p\vert^{\text{PDG}}=0.245 \,\pm 0.006$
at $95\%$ C.L.~\cite{Workman:2022ynf}. We define the BSM deviation
of helium abundance as   
\begin{equation}
\label{DeltaYpEqn}
\Delta Y_p = Y_p\vert^{\text{equation } \ref{YpEqn}}_{\text{BSM}}-Y_p\vert^{\text{equation } \ref{YpEqn}}_{\text{SM}}.
\end{equation}
The $\Delta Y_p$ contours for the Majorana and Dirac case are shown
in Figs.~\ref{bbnCnstrntPltMFig},\ref{bbnCnstrntPltDFig}, along with
overlays of the respective $\dneff$ contours in red.
In accordance with the discussion
in section~\ref{sec:constraintscurrentfuture},
a conservative upper bound is $\Delta Y_p= 0.008$ at $95\%$ C.L\@.
In both the Majorana and Dirac cases,
$\dneff$ constraints are stronger for $m_X>2m_e$.
However, in the Dirac case,
we find $\Delta Y_p \gtrsim 0.008$ in the mass region
$0.3 \; {\rm MeV} \lesssim m_X \lesssim 2 m_e$,
where the Planck constraint from the CMB, $\dneff \lesssim 0.3$
is currently weaker.  Once the future CMB observatories reach 
$\dneff \lesssim 0.15$, their constraints will exceed what can be
set by this analysis of BBN\@.


\section{Discussion}
\label{discussionSec}

In this paper we calculated the contributions
of a light $U(1)_{B-L}$ gauge boson mediator to 
$\Delta N_{\rm eff}$ as measured by the CMB
and $Y_p$ from BBN\@.
Our main result is
Figs.~\ref{dneffMFig},\ref{dneffDFig}, where we show our
predictions for $\Delta N_{\rm eff}$ as a function of
of the $U(1)_{B-L}$ gauge boson mass $m_X$ and coupling strength $g_X$,
along with Figs.~\ref{bbnCnstrntPltMFig},\ref{bbnCnstrntPltDFig}
where we show our predictions for the shift in the helium mass fraction
$\Delta Y_p$, in the same parameter space.
While substantial portions of the ($m_X$, $g_X$) parameter
space have overlapping constraints from other astrophysical
or terrestrial experiments, we find there are several regions
where future CMB observatories
\cite{SPT-3G:2014dbx,CORE:2016npo,SimonsObservatory:2018koc,NASAPICO:2019thw,Abazajian:2019eic,Sehgal:2019ewc}, 
have the opportunity to gain
sensitivity to a very weakly coupled $U(1)_{B-L}$ gauge boson
in certain mass and coupling ranges that is not accessible by
any other method.

We have also calculated $\Delta Y_p$, the helium abundance,
which serves as a proxy for $\Delta N_{\rm eff}$ as observed
by BBN\@.  This result utilized an approximation from
Ref.~\cite{Escudero:2018mvt} to estimate the change in the
helium abundance that follows from a nonzero neutrino
chemical potential, as occurs with a freeze-in abundance of a
$U(1)_{B-L}$ gauge boson.  
In figure~\ref{bbnCnstrntPltMFig}, we find that for the
Majorana case, the Planck
constraints on the CMB are generally stronger than the
BBN constraints on $Y_p$ throughout the $(m_X, g_X)$ plane.
For the Dirac case shown in figure~\ref{bbnCnstrntPltDFig}, however,
there is a nontrivial region of parameter space
(approximately $0.3 \; {\rm MeV} \lesssim m_X \lesssim 1~{\rm MeV}$,
$g_X \lesssim 10^{-10}$),
where the BBN constraints are stronger than the constraints from
$\Delta N_{\rm eff}$ from Planck data.  However, 
future CMB observatories \cite{SPT-3G:2014dbx,CORE:2016npo,SimonsObservatory:2018koc,NASAPICO:2019thw,Abazajian:2019eic,Sehgal:2019ewc}, 
will be able to fully cover
this region, and probe considerably smaller couplings,
once they achieve $\Delta N_{\rm eff} \lesssim 0.1$
sensitivity.

It is interesting to consider whether thermal equilibrium is ever achieved between $X$ and the SM\@.  In general, for large enough couplings, there will be an epoch where we can identify equilibrium.  This occurs when the comoving  energy densities of $X$ and one or more SM species reach approximately constant values.  This epoch cannot last forever, since $X$ subsequently decays back into the SM\@. For example, in figure 2 where $m_X = 10$~keV, $X$ clearly reaches
thermal equilibrium with neutrinos when $g_X = 10^{-10}$, as shown by the plateau of the energy density for a finite range of temperature.  The next contour, $g_X = 10^{-11}$, there is hardly any range of temperature where the $X$ density remains constant, and so we see that for this coupling, $X$ approaches thermal equilibrium with neutrinos but never really achieves this before $X$ subsequently decays. Similar results can also be found for the Dirac case, comparing $g = 10^{-10}$ to $g = 10^{-11}$ in figure 4.
In the region $m_X \ll 2 m_e$, broadly we find that $X$ and neutrinos approximately achieve thermal equilibrium, for a brief point in the temperature evolution of the universe, in the $\Delta N_{\rm eff} \sim 0.3$ (green) region in figure 5 and the $\Delta N_{\rm eff} \sim 0.1$ (red) region in figure 6.


We also considered semi-analytic estimates of our $\Delta N_{\rm eff}$
results.  Our focus for these estimates was the region $m_X \ll 2 m_e$,
where the only species interacting with $X$ are neutrinos.
The semi-analytic estimates, which assume thermal equilibrium is reached
between $X$ and neutrinos, allow us to obtain an
approximate value for $\Delta N_{\rm eff} \simeq 0.25$ ($0.09$) for the
Majorana (Dirac) case.
The semi-analytic estimates also permit us to determine how
our results change if there are additional degrees of freedom,
beyond the SM, that the $X$ boson is able to populate.
The physical picture is that $X$ slowly freezes in,
becomes nonrelativistic, and then decays into SM states
plus the additional degrees of freedom beyond the SM\@.
For this analysis, we considered a set $N_R$ of fermions 
carrying lepton number (right-handed neutrinos with $L = \pm 1$),
in two scenarios:  the additional fermions stay relativistic
(at least until recombination), or the additional fermions
have masses and become nonrelativistic matter.
If the fermions stay relativistic, $\Delta N_{\rm eff} \ra 0$
as $N_R$ becomes large, essentially because they dominate
the relativistic degrees of freedom after entropy equilibration.
On the other hand, if they become nonrelativistic before
recombination, they siphon off most of the relativistic
degrees of freedom into matter, and so $N_{\rm eff} \ra 0$
($\Delta N_{\rm eff} \ra -3$).  

Our results have interesting implications on other scenarios
considered in the literature:

Ref.~\cite{Krnjaic:2020znf} considered light new vectors
produced gravitationally during inflation that also couple
to neutrinos.
The coupling to neutrinos imply the inflationary produced vector  
can decay into neutrinos, giving additional contributions to
$\Delta N_{\rm eff}$, that led to constraints in the $(m_X, g_X)$ plane. 
These contributions to $\Delta N_{\rm eff}$ rely on a primordial abundance of
vector bosons from inflation, and thus constitute a
contribution that is \emph{in addition}
to the freeze-in abundance of a massive vector that couples to neutrinos.
On comparing our bounds to those from \cite{Krnjaic:2020znf}, 
we find our constraints would appear to place significant restrictions
on the scale of inflation and hence the inflationary production of
vector bosons that couple to neutrinos.  

One of the additional effects of a light $X$ boson, with a mass
well below the neutrino decoupling temperature, $m_X \ll 2$~MeV,
is that the interactions with $X$ can suppress the free streaming 
of neutrinos, and this has independent constraints from
the suppression of small scale structure 
\cite{Cyr-Racine:2013jua} that results in shifts in the phase and a decrease
in the amplitude of the acoustic peaks of the CMB
\cite{Bashinsky:2003tk, Chacko:2015noa, Baumann:2015rya, Choi:2018gho}. 
Recently, \cite{Sandner:2023ptm} analyzed the suppression of
neutrino free streaming resulting from neutrino interactions
within the context of a model with a very light $U(1)_{\mu - \tau}$
gauge boson. 
By implementing these interactions into the modeling of the CMB,
they were able to use Planck data to constrain a broad range of
$U(1)_{\mu - \tau}$ gauge boson masses between
$10^{-3}$~eV to $10^3$~eV, with the strongest bound on $g_{\mu-\tau}$
near $1$~eV, and rapidly decreasing bounds as the mass moved
away from this value.  Nevertheless, for masses $1$~eV and larger,
\cite{Sandner:2023ptm} found that the bounds from
$\Delta N_{\rm eff}$ are still stronger than those from
the suppression of the free streaming of neutrinos.
This is strongly suggestive that the bounds from
$\Delta N_{\rm eff}$ are very likely stronger than constraints
from the suppression of neutrino free streaming
in the model considered in this paper, $U(1)_{B-L}$,
though we leave a detailed
analysis of this point to future work.

Several other papers have considered a light $U(1)_{B-L}$ gauge boson's
effects in various contexts.  For example, our work complements and
extends the analysis of
\cite{Heeck:2014zfa, Abazajian:2019oqj, Luo:2020fdt, Adshead:2022ovo},
who considered bounds on a $U(1)_{B-L}$ model
in which neutrinos acquire Dirac masses.
Our work rules out substantial parts of previously allowed
parameter space in \cite{Schwemberger:2022fjl},
who considered constraints on $U(1)_{B-L}$
from neutrino-electron and neutrino-nucleon scattering.
As can be seen in figure~\ref{dneffMFig}, the remaining region
that is more strongly constrained by
neutrino-nucleon scattering occurs only when $m_X \gtrsim 10$~MeV
in the Majorana case.
By contrast, when neutrinos have Dirac masses,
the CMB constraints completely dominate the bounds
even at larger masses as shown in figure~\ref{dneffDFig}
(that includes the results from \cite{Adshead:2022ovo}).
Some groups \cite{Kaneta:2016vkq, Biswas:2016bfo,
  Kelly:2020pcy, Eijima:2022dec} considered
sterile neutrino dark matter arising from interactions
with $U(1)_{B-L}$.  While our analysis of the CMB constraints
does not rule out the main region favored by \cite{Kelly:2020pcy},
essentially all of their parameter space $m_X \lesssim 10$~MeV
would be in conflict with the CMB that extends and compliments
the existing constraints from beam dump experiments.
Ref.~\cite{Eijima:2022dec} calculated CMB constraints in the
range $0.001 \; {\rm MeV} < m_X < 1 \; {\rm MeV}$,
and while the broad trends they find in their paper are
similar with our Majorana case results, they did not include
a chemical potential for $X$ and neutrinos, making a
a quantitative comparison moot. 

Beyond $U(1)_{B-L}$, there has also been extensive discussion of
equilibration of a dark sector with neutrinos well after
BBN \cite{Berlin:2017ftj,Berlin:2018ztp,Berlin:2019pbq}.
The motivation of this work is to obtain thermal dark
matter below an MeV \cite{Berlin:2017ftj,Berlin:2018ztp}.
There are commonalities between this work and ours,
specifically the observation that if a light state couples
only to neutrinos, equilibration with neutrinos draws
heat
from the SM, and only after the species decouples, the neutrino
bath is heated, contributing to $\dneff$.
The distinction between our work and
\cite{Berlin:2017ftj,Berlin:2018ztp,Berlin:2019pbq}
is that we find the Boltzmann equations require the presence of
a nonzero chemical potential
for $X$ as well as neutrinos, in order to properly track the
evolution of the energy densities of the various species,
and therefore to accurately calculate $\dneff$.
For instance, when our $X$ boson can decay into
just one massive species (one massive ``sterile'' Majorana neutrino),
this siphons off a sufficient amount of entropy from the
left-handed neutrinos of the SM to already be excluded by
Planck data.  It would be very interesting to apply our analysis
to the specific scenarios of
\cite{Berlin:2017ftj,Berlin:2018ztp},
to determine if including a chemical potential
for the mediators and DM has an effect on the results,
but we leave for future work.

Note added:  As this work was being completed \cite{Li:2023puz} appeared
that also discussed the bounds on a model with a light
$U(1)_{B-L}$ (with Majorana neutrino masses)
and $U(1)_{\mu-\tau}$ boson from $\Delta N_{\rm eff}$.
Among the differences between their work and ours, we include
the coupling of the $U(1)_{B-L}$ mediator to charged leptons
(electrons), and the associated $1 \lra 2$, $2 \lra 2$ processes
that are essential to determining the effects 
on $\dneff$ for temperatures near and above BBN\@.

\acknowledgments

We thank P.~Asadi, M.~Dolan, M.~Escudero, and J.~Kopp for useful
discussions.  This work was supported in part by the U.S. Department
of Energy under Grant Number DE-SC0011640.


\appendix

\section{Right-handed neutrino decay in Majorana case}
\label{app:righthandedneutrinodecay}

In the regime $M_R > m_h + m_{\nu}$,
the 2-body decay is rapid, and the right-handed neutrinos will
have completely decayed well prior to BBN\@.  When $M_R < m_h$, 
the decay is 3-body, with an off-shell Higgs that
will also be suppressed by the small Yukawa coupling
$y_{\rm SM}$ of the Higgs to lighter SM fermions.
An estimate of this 3-body width to one SM fermion pair
is sufficient for our purposes,
\begin{eqnarray}
  \Gamma(\nu_R \ra \nu_L f \bar{f}) &\sim&
  \frac{y_D^2 y_{\rm SM}^2 M_R^5}{128 \pi^3 m_h^4} \, ,
\end{eqnarray}
where we have neglected the final state phase space.  
If a SM neutrino species has a mass
$m_\nu \sim (y_D v)^2/M_R$,
the width becomes
\begin{eqnarray}
  \Gamma(\nu_R \ra \nu_L f \bar{f})
  &\sim& \frac{m_\nu y_{\rm SM}^2 M_R^6}{128 \pi^3 v^2 m_h^4} 
\label{eq:widthwithneutrinomass}
\end{eqnarray}
leading to a lifetime
\begin{eqnarray}
  \tau(\nu_R \ra \nu_L f \bar{f}) &\sim& (1 \; {\rm s})
  \times \left( \frac{10 \; {\rm GeV}}{M_R} \right)^6
  \times \left( \frac{0.1 \; {\rm eV}}{m_\nu} \right)
  \times \left( \frac{y_b}{y_{\rm SM}} \right)^2 \, , 
\end{eqnarray}
where we have normalized the Yukawa coupling $y_{\rm SM}$
to the $b$-quark Yukawa coupling.  If $M_R \gtrsim 2 m_b$,
we expect $\nu_R$ to decay predominantly a $b\bar{b}$ pair
since this is the heaviest SM fermions that are kinematically
available for the 2-body decay.  If $M_R$ is slightly less
than the $2 m_b$ threshold, decays to $c\bar{c}$ and $\tau^+\tau^-$
would be present, with $y_{\rm SM}$ slightly smaller than $y_b$.

In any case, the above estimate demonstrates that for
$M_R \lesssim 10$~GeV, $\nu_R$ are comparatively long-lived and
can disrupt BBN light element abundance predictions due to the
electromagnetic energy deposition \cite{Jedamzik:2006xz}. 
Given that the width, equation~(\ref{eq:widthwithneutrinomass}),
depends on the 6th power of $M_R$, one only needs slightly
larger masses, $M_R \gtrsim 20$~GeV, to cause $\nu_R$ to decay
sufficiently fast to completely avoid BBN constraints.

This leads to two possibilities in the
Majorana scenario:

If $\nu_R$ was in thermal equilibrium in the early Universe,
$M_R \gtrsim 20$~GeV is required to avoid BBN constraints,
and thus $v_X \gtrsim (20 \; {\rm GeV})/y_M$.
If $X^\mu$ acquires \emph{all} of its mass
from just the $\phi_X$ vev, then $m_X$ is determined once $g_X$
is specified, specifically, $m_X = 2 g_X v_X$.
The scalar Higgs sector of $U(1)_{B-L}$ need not be minimal.
There could be other scalars with $B-L$ charge that differ
$\phi_X$, i.e., $\phi_{X'}$ with $q_{X'} \not= \pm 2$.
These do not lead to contributions to the Majorana masses,
but will lead to an additional contribution to the $X^\mu$
gauge boson mass, $m_X^2 = g_X^2 \left( 4 v_X^2 + q_{X'}^2 v_{X'}^2 \right)$.
Hence, we see that the minimal case $m_X = 2 g_X v_X$ is a
\emph{lower} bound on $m_X$ for a generic $U(1)_{B-L}$
breaking sector.  Said directly, the bound 
$M_R \gtrsim 20$~GeV implies
\begin{eqnarray}
  g_X &<& \frac{m_X}{2 v_X} \nonumber \\
  &\lesssim& 2.5 \times 10^{-5} \frac{m_X}{1 \; {\rm MeV}}
          \qquad \mbox{(if $\nu_R$ in early thermal equilibrium).} 
\label{eq:gXbound}
\end{eqnarray}

On the other hand, it could be that $\nu_R$ was never in thermal
equilibrium in the early universe.  In this case, no population of
$\nu_R$'s were generated, and there is no bound on
$g_X$ from $\nu_R$ decay.


\section{Finite Temperature Effects}
\label{app:finiteTemperatureEffects}

In the SM, finite temperature effects \cite{Mangano:2001iu, Fornengo:1997wa, Escudero:2018mvt} result in an $\mathcal{O}(0.01)$ correction to $\neff$. One example is the finite shift in the photon's propagator due its self-energy $\Pi_{\mu\nu}$ in presence of an $\ee$ background. This shift generates a complex effective mass for the photon and modifies the dispersion relation of the longitudinal and transverse spin polarizations of the photon. To $\mathcal{O}(\alpha)$, a simplified version of the effective masses is
 \begin{equation}
\label{effectiveMassesOfPhoton}
m_\gamma^2 =
\begin{cases}
4\pi\alpha (n_e/m_e) & (T\ll m_e) \\
\frac{2}{3}\pi\alpha T^2 & (T\gg m_e).
\end{cases}
\end{equation} 
A similar story generates an effective mass for electrons
\begin{equation}
\label{effectiveMassesOfElectron}
m_{e,\text{eff}}^2(T) \approx m_{e,0}^2+\pi \alpha T^2.
\end{equation} 
Note that $m_\gamma(T)<2 m_{e,\text{eff}}(T)$ and hence photon decay to $\ee$ pair is always kinematically forbidden.

Finite temperature correction to early universe dynamics with an additional hidden photon (which we label $X$ for convenience) has also been explored in \cite{Jaeckel:2008fi, Mirizzi:2009iz, Arias:2012az, McDermott:2019lch, Witte:2020rvb, Caputo:2020rnx, Caputo:2020bdy} and most relevant to our discussion here are \cite{Redondo:2008ec, Fradette:2014sza, Li:2020roy, Li:2023vpv}. We will summarize this discussion and use the results to calculate the contribution of finite temperature effects in our model. In what follows, we only take the transverse part of photon polarization tensor, since it was shown in \cite{Fradette:2014sza} that the longitudinal part of the photon polarization tensor does not significantly contribute to the finite temperature effects.

Following \cite{Li:2023vpv}, for a given interaction $\aab \rightarrow X$, one can write the additional interaction $\aab \rightarrow \gamma \rightarrow X$, where $\gamma \rightarrow X$ is the contribution of all allowed fermion loops in medium. In vacuum, such interactions are loop-suppressed and are negligible. However, in a medium with finite temperature/density, these interactions are important and can, in some cases, dominate the rate \cite{Redondo:2008aa, An:2013yfc,  Redondo:2013lna, Chang:2016ntp, Hardy:2016kme, Li:2023vpv}. The  $\aab \rightarrow \gamma \rightarrow X$ diagram is proportional to
\begin{equation}
\label{inMediumCorrectionMixing}
e \frac{-i}{k^2-\Pi_{\gamma-(f) -\gamma}}i\Pi_{\gamma -(f)-X},
\end{equation}
where $e$ is the electromagnetic coupling, the second piece is the in-medium photon propagator, and the third is the induced $\gamma-X$ mixing in medium. 
Assuming the two polarization tensors have the same fermion loops, then $\Pi_{\gamma -(f)-X} = (g_X/e) \Pi_{\gamma -(f)-\gamma}$. The sum of the two diagrams is now given by 
 \begin{align}
\label{sumOfVacAndMedInteractions}
(\aab \rightarrow X)+(\aab \rightarrow\gamma \rightarrow X) &\propto g_X\left(1+  \frac{ \Pi_{\gamma \gamma}}{k^2-\Pi_{\gamma\gamma}}\right)\\
&=g_X \frac{m_X^2}{m_X^2-\Pi_{\gamma\gamma}},
\end{align} 
and the modification to the coupling in medium can be written as,
\begin{equation}
\label{effectiveCouplingResponse}
g_{X,m}^2=\frac{g_X^2 m_X^4}{| m_X^2-\Pi|^2} = \frac{g_X^2 m_X^4}{(m_X^2 -m_\gamma^2)^2+(\text{Im} \, \Pi )^2},
\end{equation}
where $\text{Re}\,\Pi =m_\gamma^2$ in Eq.~\ref{effectiveMassesOfPhoton}.
The imaginary part $\text{Im}\,\Pi$ \cite{Weldon:1983jn} encodes the rate at which the photon's distribution approaches a thermal equilibrium,
\begin{equation}
\label{ImPolarizationResponse}
\text{Im} \, \Pi = - \omega \Gamma \equiv -\omega[\Gamma_{\text{decay}}(\omega) +\Gamma_{\text{inverse}}(\omega)].
\end{equation}
The imaginary part of the self-energy arises at two-loops and we generically have $\text{Im}\,\Pi \ll \text{Re}\,\Pi$. The effective coupling in Eq.~\ref{effectiveCouplingResponse} can be discussed in three regimes;  $m_\gamma(T) \gg m_X$, $m_X = m_\gamma(T_r)$, and $m_\gamma(T) \ll m_X$. The middle regime, $m_X = m_\gamma(T_r)$, is resonant. 
If the resonance occurs when the electrons are relativistic, $T\gg m_e$, the resonance temperature is
\begin{equation}
\label{resonanceCondition}
T_r = \sqrt{\frac{3}{2\pi \alpha}}m_X \approx 8m_X \;\;\;\;\;\;  (T\gg m_e).
\end{equation}
If the electrons are non-relativistic when the resonance occurs then $m_\gamma(T_r) \propto T_r^{3/2}e^{-m_e/T_r}$ as per Eq.~\ref{effectiveMassesOfPhoton}. The photon plasma mass is extremely sensitive to $T$ and the resonance occurs around $T_r \sim 0.2 m_e$ for a wide range of masses $m_X \sim(1-10^5 \,\text{eV})$. We now discuss the regimes.

\subsection{$m_\gamma(T) \ll m_X$}

In this regime, the in-medium coupling $g_{X,m}$ reduces to its vacuum value $g_{X}$.
We start by examining the entropy scaled number density $Y_X=n_X/s$ produced from forward interactions to $X$ using only the vacuum coupling. We write the simplest Boltzmann equation 
\begin{equation}
\label{YXboltzmannEqn}
s\dot{Y}_X=\dot{n}_X+3Hn_X= \sum_{a}\frac{3}{2\pi^2}\Gamma_{X\rightarrow a \bar{a}}m_X^2 T K_1(m_X/T).
\end{equation}
Here, we have assumed massless particles $a$ with zero chemical potential and took the Maxwell-Boltzmann approximation for distributions in the collision term. Explicitly, the collision term includes $\ee \rightarrow X$ and $\nunui \rightarrow X$. The final $Y_X^f$ from the following two forward interactions can be calculated 
\begin{equation}
\label{YXFinal}
Y_X=\int_0^\infty dT \frac{\dot{Y}_X}{H(T) T}=\frac{3}{16\pi^2}\frac{g_X^2 m_X^4 }{[Hs]_{T=m_X}} \times
\begin{cases}
(1) &(e^+e^- \rightarrow X)\\
(\frac{3}{2})& (\nu\bar{\nu} \rightarrow X) \, . 
\end{cases}
\end{equation}
In the next sections, we show that the finite temperature effects contributions to $Y_X$ are negligible compared to Eq.~\ref{YXFinal}.

\subsection{$m_\gamma(T) \gg m_X$}

Here, the coupling $g_{X,m}$ is suppressed by a factor of $m_X^2/m_\gamma^2$ compared to its vacuum counterpart. However, this suppression occurs above resonance, $T_r>8m_X$, where the $X$ production rates are inefficient even in the vacuum case. In our Boltzmann evolution of $X$ population and for small values of $g_X$, a negligible amount of $X$ is produced in this temperature range. This makes this regime unimportant for our analysis, since this suppression would only decrease the amount of $X$ produced. We ignore this regime in all subsequent discussion.

\subsection{$m_\gamma(T) = m_X$ resonance}


We now discuss the interactions that are allowed during the resonance and thus contribute to the resonant $X$ production. \emph{Importantly, the second diagram in Eq.~\ref{sumOfVacAndMedInteractions} does not exist for $\nunui \rightarrow X$. The neutrino coalescence rate is not changed by finite temperature effects}  \footnote{The transfer rate $\delta \rho_{X\rightarrow \nunui}/\delta t$ is proportional to the distribution $f_X$. The only modification to this rate at a given temperature stems from modifications to $f_X$ from other enhanced electron interactions. Nonetheless, The coupling $g_X$ in the neutrino coalescence rate is unchanged.}. Thus, Our treatment of neutrinos rates is complete and neutrinos do \emph{not} contribute to the resonance. Furthermore, $\ee \rightarrow X$ is kinematically allowed only for $m_X>2m_{e,\text{eff}}$ or when 
\begin{equation}
\label{coalescenceCondition}
T\leq T_c= \sqrt{\frac{m_X^2-4m_{e,0}^2}{4\pi \alpha}} < T_r. 
\end{equation}
Hence, $\ee$ coalescence also does \emph{not} contribute to the resonance since the interaction is only allowed for temperatures strictly below $T_\text{res}$. This point is emphasized in \cite{Redondo:2008ec}, but missed in more recent literature \cite{Fradette:2014sza, Li:2020roy}. The contribution to the resonance in both our and the hidden photon case, comes from pair annihilation $\ee \lra \gamma X$ and semi-Compton $e \gamma \lra e X$.
The resonant production increases the initial value of $\rho_X$ at the expense of energy density of the $e-\gamma$ plasma. The decay of $X$ back into SM particles is unchanged. Explicitly, a larger initial $\rho_X$ or a smaller initial $\rho_{\gamma-e}$ causes an increase the value of $\dneff$.


In this case, the entropy scaled number density $Y_{X, \,\text{res}}$
produced from the resonance through the interaction $a+b \rightarrow X + c $
is \cite{Redondo:2008ec},
\begin{equation}
\label{YXboltzmannEqn2To2}
s\dot{Y}_X =\int dn_a dn_b \tilde{\sigma}_{2\rightarrow 2}(s) v_{\text{M\"{o}l}}\frac{g_X^2 m_X^4}{(m_X^2 -m_\gamma^2)^2+(\text{Im}\Pi )^2},
\end{equation}
where $\int dn_i=\int g_i dp_i^3f_i/(2\pi)^3$, the $\tilde{\sigma}=\sigma/g_{X,m}^2$, and $v_{\text{M\"{o}l}}$ is the Moeller velocity. At resonance, expanding $m_\gamma^2=m_X+(m_\gamma^2)' (T-T_r) + \dots$ and using the narrow width approximation, 
\begin{align}
\label{couplingJT}
\frac{g_X^2 m_X^4}{(m_X^2 -m_\gamma^2)^2+(\text{Im}\Pi )^2}&= g_X^2 m_X^4\frac{\pi }{(m_\gamma^2)'  \text{Im}\Pi} \delta(T-T_r)\nonumber\\
&= g_X^2 m_X^2\frac{T}{j(T)}\frac{\pi }{\text{Im}\Pi} \delta(T-T_r),
\end{align}
where $j(T) \equiv \frac{T}{m_\gamma^2}(m_\gamma^2)'$ is a function defined in \cite{Redondo:2008ec}. The integrals can be performed to obtain
\begin{align}
\label{YXboltzmannResonance}
Y_{X,\, \text{res}} &= \frac{2}{3\pi}g_X^2 m_X^2\int_{m_X}^{\infty} d\omega \frac{1}{j(T_r)[Hs]_{T=T_r}}\frac{\sqrt{\omega^2-m_X^2}}{e^{\omega/T_r}-1}\nonumber\\
&\approx \left(\frac{\pi}{9 j(T_r)}\frac{m_X^3}{T_r^3} \right) \left(
\frac{[\sqrt{g_{*}}g_{*S}]_{T=m_X}}{[\sqrt{g_{*}}g_{*S}]_{T=T_r}}\right)
\frac{g_X^2 m_X^4 }{[Hs]_{T=m_X}}
 \equiv \mathcal{A} \mathcal{B} \frac{g_X^2 m_X^4 }{[Hs]_{T=m_X}},
\end{align}
where $g_{*},g_{*S}$ are the standard relativistic degrees of freedom in energy and entropy density.
The factors $\mathcal{A}$  and $\mathcal{B}$ are defined as the terms in first and second set of parentheses respectively. In all cases, $T_r>m_X$ and $\mathcal{B} \leq 1$ and $\mathcal{B}$ thus can only decrease the prefactor in Eq.~\ref{YXboltzmannResonance} in comparison to Eq.~\ref{YXFinal}. We will conservatively ignore $\mathcal{B}$ in our argument as it depends on value of $m_X$. Furthermore, In a more detailed treatment of this derivation, $m_\gamma^2$, and subsequently $T_r$ and $j(T)$,  have $\omega$ dependence. However, a numerical evaluation of the integral in Eq.~\ref{YXboltzmannResonance} taking this dependence into accounts does not significantly change the values of $Y_{X,\text{res}}$ discussed here.

\subsection{Implications of the resonance on our results}

For $m_X \gtrsim 2 m_e$,  the resonance occurs at $T_r\sim8m_X$ where $j(T_r)=2$. With these substitutions, the factor $\mathcal{A}$ in Eq.~\ref{YXboltzmannResonance} is suppressed by $1/8^{3}$ and is $\mathcal{A}\sim \mathcal{O}(10^{-4})$. Since $e^+e^- \ra X$ is active, comparing these values to $\mathcal{O}(10^{-2})$ prefactor in Eq.~\ref{YXFinal} one finds that finite temperature effects contribute at most a few percent to $Y_X$.

For  $m_X\ll m_e$, the resonance occurs around $30-200\,\kev$ for a wide range of masses $m_X \sim 1-10^4\,\text{eV}$. Importantly, the resonance is the only active mechanism that permits energy from the $e^+e^-\gamma$ plasma to be transferred into $X$ bosons. For example, for $m_X=10\,\kev$, the resonance temperature is $T_r \approx 200\,\kev$ and $j(T_r)\approx 4$ giving $\mathcal{A}\sim\mathcal{O}(10^{-5})$. For $m_X=10\,\text{eV}$, then $T_r \approx 40\,\kev$ and $j(T_r)\approx 10$ giving $\mathcal{A} \sim \mathcal{O}(10^{-13})$.  Consequently, when the  resonance process dominates the contribution to $\dneff$, the constraints on $g_X$ become increasingly weaker at lower $m_X$ mass, illustrated by the $\dneff = 0.4,0.5$ contours in figure~\ref{dneffMFig} and $\dneff = 0.2$ and larger contours in figure~\ref{dneffDFig}.

\bibliographystyle{JHEP}
\bibliography{NeffJCAP}

\end{document}